\documentclass[prl, superscriptaddress,twocolumn,10pt,
 amsmath,
 amssymb,
 aps,
]{revtex4-2}

\usepackage{graphicx}

\usepackage{bm}
\usepackage[dvipsnames]{xcolor}
\usepackage{subfigure}

\usepackage{hyperref}

\definecolor{darkpink}{rgb}{0.91, 0.33, 0.5}

\begin{document}

\title{Quantum optics measurement scheme for quantum geometry \\ and topological invariants}%

\author{Markus Lysne}
\affiliation{Department of Physics, University of Fribourg, CH-1700 Fribourg, Switzerland}
\author{Michael Sch\"uler}
\affiliation{Department of Physics, University of Fribourg, CH-1700 Fribourg, Switzerland}
\affiliation{Laboratory for Materials Simulations, Paul Scherrer Institute, CH-5232 Villigen PSI, Switzerland}
\author{Philipp Werner}
\affiliation{Department of Physics, University of Fribourg, CH-1700 Fribourg, Switzerland}

\date{\today}

\begin{abstract}
 We show how a quantum optical measurement scheme based on heterodyne detection can be used to explore geometrical and topological properties of condensed matter systems. Considering a 2D material placed in a cavity with a coupling to the environment, we compute correlation functions of the photons exiting the cavity and relate them to the hybrid light-matter state within the cavity. Different polarizations of the intracavity field give access to all components of the quantum geometric tensor  
on contours in the Brillouin zone defined by the transition energy. Combining recent results based on the metric-curvature correspondence with the measured quantum metric allows us to characterize the topological phase of the material. Moreover, in systems where $S_z$ is a good quantum number, the procedure also allows us to extract the spin Chern number. As an interesting application, we consider a minimal model for twisted bilayer graphene at the magic angle, and discuss the feasibility of extracting the Euler number.  

\end{abstract}

\maketitle

{\it Introduction.} 
Geometrical and topological properties of Bloch states play an important role in modern condensed matter physics \cite{Qi_2008, Hasan_2010}. A prominent manifestation of %this 
a global topological property 
is the quantized value of the Hall conductivity, as obtained by linear response calculations \cite{Klitzing_1980, Haldane_1988}. Quantum geometry, on the other hand, refers to quantities that are local in momentum space, such as the Berry curvature and quantum metric \cite{Mera_2022}. They are related to topology, but also influence the motion of electrons in the Brillouin zone (BZ)
\cite{Lapa_2019, Leblanc_2021}, non-linear optical responses \cite{Ma_2021, Ahn_2020, Ahn_2022, Orenstein_2021, Morimoto_2016,Lysne_2021} and flat-band superconductivity \cite{Peri_2021, Peotta_2015}. 

Since the relation between geometry, topology and observable quantities is often not obvious, it is important to identify experimental probes which allow to measure these quantities, or at least provide relevant bounds \cite{Gersdorff_2021}. To this end, several schemes utilizing linear response for the measurement of quantum geometry have been proposed \cite{Ding_2022, Ozawa_2018, Gersdorff_2021, Asteria_2019, Tran_2017}. However, while the spectroscopy of topological states of matter  using semiclassical descriptions has been studied extensively \cite{Tran_2017,schuler_tracing_2017,Asteria_2019,Schuler_2020,baykusheva_all-optical_2021,chen_optical_2022}, the quantum optics side is less explored. 
In quantum optics, entanglement between light and matter can lead to novel phenomena \cite{Sentef_2020, Ruggenthaler_2018, Sentef_2018, Schlawin_2022}, and it can imprint properties of the matter system into the photon field. Since current-current correlation functions are fundamentally linked to the quantum metric \cite{Neupert_2013, Kashihara_2022}, this suggests that the study of photon correlation functions of a cavity system provides a potentially fruitful avenue for probing a material's geometrical and topological properties.

Here, we propose a quantum optical measurement scheme based on heterodyne detection \cite{Grynberg_2010, Vogel_2006}. The idea is to access general photon correlation functions inside a cavity, enabled by a coupling between the cavity and the environment, 
while 
a second photon field is superimposed on the field emitted from the cavity in order to slow down the time dependence of the signal. We demonstrate that such correlation functions can be directly related to the quantum geometric tensor, a quantity which encompasses both the Berry curvature and quantum metric. The latter is related to topology via the localization dichotomy \cite{Ozawa_2021, Marzari_1997, Mera_2022}. Lastly, our method also provides an energy resolution which allows us to devise useful bounds on topological invariants of 2D systems. 

\begin{figure}[b]
	\includegraphics[width=\columnwidth]{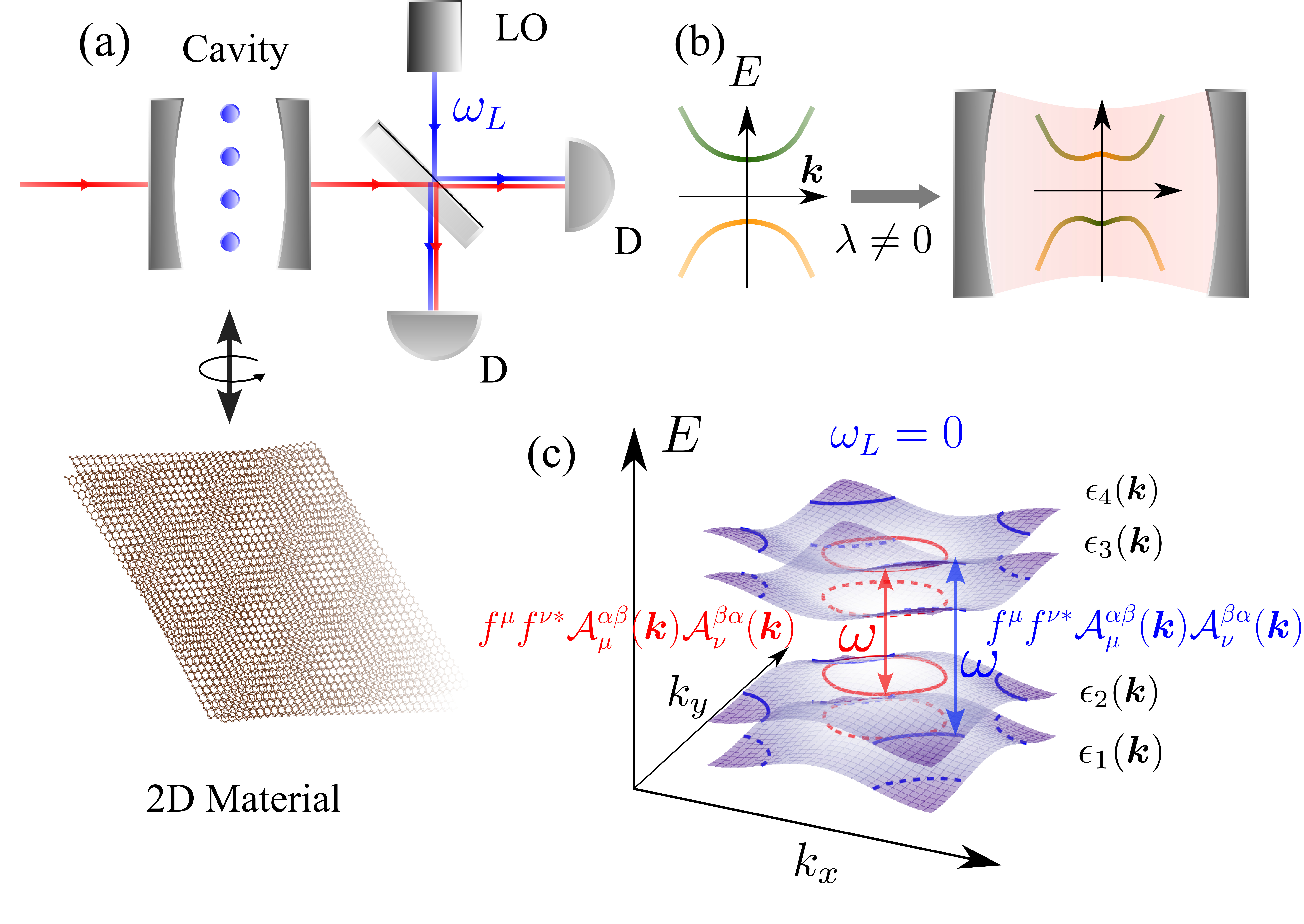}
\vspace{-10mm}
  \caption{ (a) Proposed heterodyne detection setup with the cavity depicted on the left. The electric field exiting the cavity (red) impinges on a beam splitter along with a coherent laser source (blue) to produce a superimposed signal which is detected at two photodetectors (right and bottom right). (b) Depiction of the hybrid light-matter state arising from the cavity light-matter coupling (parameter $\lambda$). (c) $k$-space contours in a multi-band model defined by a fixed excitation energy $\omega$.
  }
  \label{fig:setup}
\end{figure}

{\it Setup.}
The heterodyne detection setup is sketched in Fig.~\ref{fig:setup}(a). A Fabry-Perot cavity containing a 2D material is coupled to the environment, enabling an electric field to be transmitted through it while picking up signatures of the hybrid light-matter state within the cavity \cite{Walls_2007} (Fig.~\ref{fig:setup}(b)). By placing a 50:50 beam splitter 
behind the output port 
and superimposing a second \emph{coherent} laser source - henceforth referred to as local oscillator (LO) - it is possible to slow down the signal emitted from the cavity. 
This allows to bypass limitations in the time resolution of photodetectors, which is the goal of heterodyne detection \cite{Vogel_1995, Vogel_2006, Grynberg_2010}, and enables the measurement of various photon correlation functions in the two detectors denoted by ``D" in Fig.~\ref{fig:setup}(a) \cite{Vogel_2006, Vogel_1995, Welsch_1999}.

We begin by describing the intra-cavity Hamiltonian which can be decomposed as $\hat{H}_{\text{cav}} = \hat{H}_{\text{free}} + \hat{H}_{I} + \hat{H}_{\text{mat}}$. The free cavity field is described by $\hat{H}_{\text{free}}= \hbar  \Omega { \hat{ a} }_{ }^{ \dagger } { \hat{ a} }$, with $\hat a$ the photon annihilation operator and $\Omega$ the cavity frequency, while $\hat{H}_{\text{mat}}$ refers to the Hamiltonian of the electron system. The light-matter coupling in the Coulomb gauge and in the single-mode approximation reads 
\begin{equation}
  \begin{aligned}
  \hat{ H}_{ I}=& - \frac{ q}{ m} \hat{\boldsymbol{ A}} \cdot \sum^{}_{\boldsymbol{k}, \alpha\beta} \hat{ c}_{ \boldsymbol{k}, \alpha}^{ \dagger }  \langle \psi_{ \boldsymbol{k}, \alpha} | \hat{\boldsymbol{p}} | \psi_{ \boldsymbol{k},\beta} \rangle  \hat{ c}_{ \boldsymbol{k}, \beta}  \\
  &
  + \frac{ q^2}{ 2 m} { \hat{ A}_{ } }^{ 2} \sum_{ \boldsymbol{k}, \alpha} { \hat{ c} }_{ \boldsymbol{k}, \alpha}^{ \dagger } { \hat{ c} }_{ \boldsymbol{k}, \alpha},
\end{aligned}
\end{equation}
where $q$ and $m$ represent the electron's charge and mass, respectively, $\hat{\boldsymbol{A}} = \lambda[\boldsymbol{f} \hat{a}+ \boldsymbol{f}^* \hat{a}^\dagger]$ the vector potential treated within the dipole approximation, $\lambda$ the light-matter coupling parameter and $\boldsymbol{f}=( f^x, f^y )$ the mode function of the vector potential. $ \hat{c}_{\boldsymbol{k}, \alpha}^\dagger$ is the creation operator of an electron in the Bloch state $ | \psi_{\boldsymbol{k},\alpha} \rangle$ and the material is represented in this basis by a non-interacting $N$-band model $\hat{H}_{\text{mat}} = \sum_{\boldsymbol{k}}\sum_{\alpha=1}^N \epsilon_{\alpha}(\boldsymbol{k})  \hat{c}_{\boldsymbol{k},\alpha}^\dagger \hat{c}_{\boldsymbol{k},\alpha}$ with $M$ occupied bands. $| u_{\boldsymbol{k},\alpha} \rangle=e^{-i \boldsymbol{k} \cdot \hat{\boldsymbol{r}}} | \psi_{ \boldsymbol{k},\alpha} \rangle$ and $\epsilon_\alpha (\boldsymbol{k})$ are solutions to the eigenvalue equation $\hat{\mathcal{H}}(\boldsymbol{k}) | u_{\boldsymbol{k}, \alpha} \rangle = \epsilon_\alpha (\boldsymbol{k}) | u_{\boldsymbol{k},\alpha }\rangle$. The light-matter coupling in the cavity gives rise to a hybrid matter-photon state, which manifests itself in the correlator of the intra-cavity photon mode,
\begin{align}
    \label{eq:final_corr}
    & \langle { \hat{ a} }^{ \dagger } (t) \hat{ a} (t') \rangle - \langle  \hat{ a}^{ \dagger } (t) \rangle \langle  \hat{ a} (t') \rangle 
    \approx \Big( \frac{ q\lambda}{ \hbar } \Big)^2 \sum^{}_{\boldsymbol{k}} \sum_{\alpha=1}^{M} \nonumber\\
    &\quad \times \sum_{\beta=M+1}^{N} e^{  i \epsilon_{\beta\alpha}( \boldsymbol{k}) t_{\text{rel}}} { f}^{ \mu} { f}^{ \nu *} \mathcal{A}^{ \alpha, \beta}_\mu (\boldsymbol{k}) \mathcal{A}^{ \beta, \alpha}_\nu(\boldsymbol{k}),
\end{align}
where $\langle \cdots \rangle$ is computed over the density matrix of the cavity according to the Gell-Mann low theorem \cite{Stefanucci_2013} by adiabatically switching on $\hat{H}_{I}$ and $ \epsilon_{ \beta\alpha}(\boldsymbol{k}) \equiv \tfrac{1}{\hbar}( { \epsilon}_{ \beta} (\boldsymbol{k}) - { \epsilon}_{ \alpha} (\boldsymbol{k}) )$ is the $\boldsymbol{k}$-dependent valence-conduction band-gap. $t_{\text{rel}} \equiv t-t' $ and $ \mathcal{A}_{ \alpha \beta}^{ \mu}( \boldsymbol{k} ) = \langle u_{\boldsymbol{k} ,\alpha} | i \partial_{\mu} | u_{\boldsymbol{k} ,\beta} \rangle$ (with $\partial_\mu = \partial/\partial k^\mu $) the non-Abelian Berry connection.
 We use the Einstein summation convention for the spatial indices $\mu,\nu$. Equation~\eqref{eq:final_corr} neglects a higher-order term in $\lambda$ and is valid in the regime $\Omega \ll  \min_{\boldsymbol{k} \in \text{BZ}} |\epsilon_{\alpha,\beta}(\boldsymbol{k}) |$, i.e. away from topological transitions (see Supplemental Material (SM)). 

Our goal is to connect the intra-cavity correlator~\eqref{eq:final_corr} to the detected photons in the setup of Fig.~\ref{fig:setup}(a). To this end we employ the theory of photodetection \cite{Vogel_2006} and input-output theory \cite{Viviescas_2003}, which yields
\begin{align} 
  \label{eq:main_eq}
	   &\frac{ \overline{ n(t, \Delta t) n(t', \Delta t)  } - \overline{n(t,\Delta t) } \cdot  \overline{ n(t', \Delta t) } }{ \overline{ n(t, \Delta t)}}
	    \approx \nonumber\\
	    &\mathcal{D} \sum^{}_{\boldsymbol{k}} \sum_{\alpha=1}^{M}\sum_{\beta=M+1}^{N}  \big[ e^{ i ( \epsilon_{\beta\alpha}(\boldsymbol{k}) -  \omega_L ) t_\text{rel}}  +\text{H.c.} \big] \nonumber\\
	    &\hspace{26mm}\times  
	    { f}^{ \mu} { f}^{ \nu *} \mathcal{A}^{ \alpha, \beta}_\mu(\boldsymbol{k}) \mathcal{A}^{ \beta, \alpha}_\nu(\boldsymbol{k})  
\end{align}
for the correlations between the photon counts $n(t,\Delta t)$ %at time $t$ 
at a single detector, see SM. %at times $t$ ($t'$)
The frequency of the local oscillator is $\omega_L$, and coefficients related to the input-output theory and beam-splitter relations are subsumed into the coefficient $\mathcal{D}$. \

Equation~\eqref{eq:main_eq} provides a direct link to the non-Abelian quantum geometric tensor (QGT),
\begin{equation} \label{eq:QGT}
\begin{aligned}
	Q_{\mu\nu}^{\alpha\beta}( \boldsymbol{k}) &= \sum^{N}_{\gamma=M+1} \mathcal{A}^{ \alpha \gamma}_{ \mu}( \boldsymbol{k} ) \mathcal{A}^{ \gamma \beta}_{ \nu}( \boldsymbol{k} ) 
	= g_{\mu\nu}^{\alpha\beta}(\boldsymbol{k}) - \frac{i}{2}\mathcal{F}_{\mu\nu}^{\alpha\beta}(\boldsymbol{k}),
\end{aligned}
\end{equation}
which can be decomposed into the quantum metric ($g_{\mu\nu}^{\alpha\beta}(\boldsymbol{k})$) and Berry curvature ($\mathcal{F}_{\mu\nu}^{\alpha\beta}(\boldsymbol{k})$) contributions. In these expressions, $\mu,\nu=x,y$ and $\alpha,\beta\in[1,M] $.   
Since ${ f}^{ \mu} {f}^{\nu  *} \sum^{}_{ \boldsymbol{k}}  \textrm{Tr}_b[{ Q}_{ \mu\nu}(\boldsymbol{k})]$ with $\textrm{Tr}_b[ \dots]=\sum_{\alpha=1}^M [\dots]$ can be measured by Eq.~\eqref{eq:main_eq}, we gain access to (i) the Chern number if ${\boldsymbol{f}}_{ \circlearrowright (\circlearrowleft)} = ( 1 , \pm i )$, 
(ii) general diagonal components of $Q$ if ${\boldsymbol{f}} = (1,0) $ or $(0,1)$,
and (iii) off-diagonal elements of $g$ if ${\boldsymbol{f}}_{ \pm} = ( 1, \pm 1)$ -- in a way analogous to interband transitions driven by classical light~\cite{tran_probing_2017,schuler_tracing_2017,Ozawa_2019,Gersdorff_2021}.

For illustrative purposes, let us consider ${ \omega}_{ L} =0$. Since $ \epsilon_{ \beta\alpha}(\boldsymbol{k}) >0$, a Fourier transform of the first term on the right hand side of Eq.~\eqref{eq:main_eq} would select contours in the BZ with fixed energy difference $ \omega$ and sum up the values of $ f^{\mu} { f}^{ \nu *} \mathcal{A}^{ \alpha, \beta}_{ \mu} ( \boldsymbol{k} ) \mathcal{A}^{ \beta, \alpha}_{ \nu}( \boldsymbol{k} ) $ at the corresponding $\boldsymbol{k}$-points. 
This is shown in Fig.~\ref{fig:setup}(c) for two different $\omega$ represented by red and blue colors. 
 The role of $ { \omega}_{ L}$ is to lower the frequencies at which these resonances occur, consistent with the goal of heterodyne detection \cite{Grynberg_2010}. 

\begin{figure*}[t]
  \includegraphics[width=\textwidth]{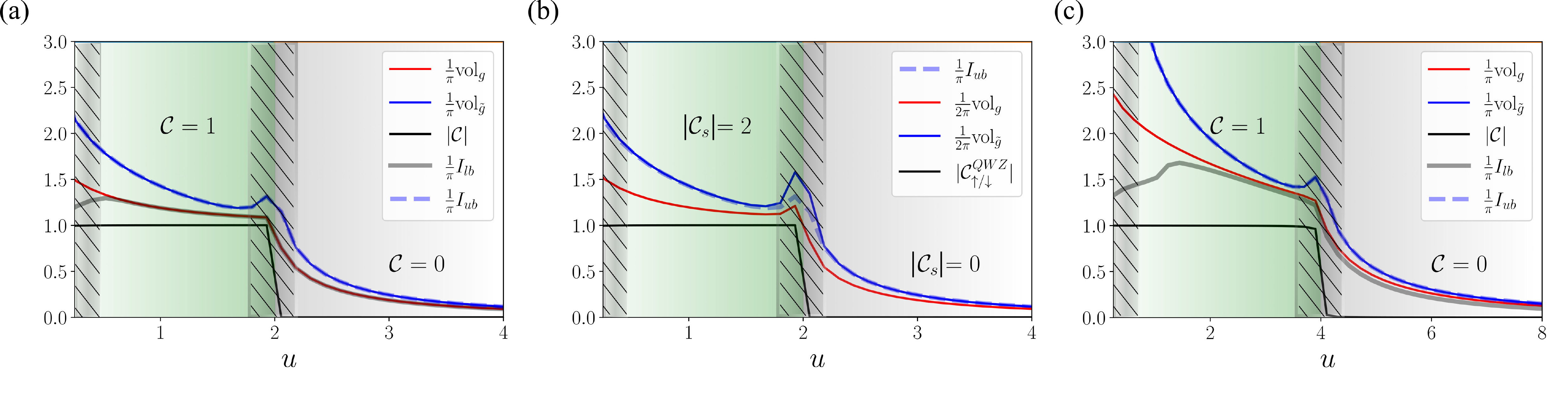}
  \caption{ Chern numbers and various bounds for (a) the QWZ model and (b) the SOC model with $\Delta=0.1$  
  and (c) the result of the three band model. All quantities of Eq.~\eqref{eq:ineqs} are plotted as a function of the parameter $u$ which controls the topological state of the models. The hashed regions near the topological transitions roughly indicate the range of $u$ where Eq.~\eqref{eq:main_eq} is not valid. For computing $I_\text{lb}$ and $I_\text{ub}$ we approximate $ \delta(x)$ by $\tfrac{1}{\sqrt{\pi}|a|} \text{exp}\{ - (x/a)^2\}$ with $a=0.1$.}   
  \label{fig:inequalities}
\end{figure*}

{\it Band topology and localization dichotomy.}
The positive semi-definite nature of $Q_{\mu\nu}(\boldsymbol{k})$ implies certain inequalities involving the quantum metric and topological invariants \cite{Ozawa_2021, Xie_2020}. The inability to devise a smooth gauge of Bloch functions in a non-trivial topological phase is an obstruction to creating maximally localized Wannier functions \cite{Marzari_2012}. Since the Wannier spread is directly related to the quantum metric \cite{Marzari_1997}, we will utilize this localization dichotomy \cite{Monaco_2018} to relate Eq.~\eqref{eq:main_eq} to topological invariants. For Chern insulators in 2D, the localization dichotomy manifests itself in bounds for the Chern number, $\mathcal{C}= \tfrac{1}{2\pi} \int d^2 \boldsymbol{k} \text{Tr}_b [\mathcal{F}_{xy}(\boldsymbol{k})]$ \cite{Ozawa_2021}\footnote{We use integrals whenever discussing topological invariants from now on, but leave the discrete sums in Eq.~\eqref{eq:final_corr} to enable a finite system description.}, as  
%$\mathcal{C}$:}
\begin{equation} \label{eq:metric_curvature}
  \pi | \mathcal{C}| \leq \textrm{vol}_g \leq \textrm{vol}_{\tilde{g}},
\end{equation}
where the so-called complexity of the band $\textrm{vol}_g = \int_{ }^{ } d^2 \boldsymbol{k} \sqrt{ \textrm{det}( \textrm{Tr}_b [ g(\boldsymbol{k})] ) }$ measures the BZ area with respect to the quantum metric \cite{Marzari_1997} and $\textrm{vol}_{\tilde{g}} = \sqrt{ \int \! d^2{\boldsymbol{k}} \textrm{Tr}_b[ { g}_{ xx}(\boldsymbol{k})  ]  \int \! d^2{\boldsymbol{k}} \textrm{Tr}_b[ { g}_{ yy}(\boldsymbol{k})  ]
- (  \int \! d^2{\boldsymbol{k}} \textrm{Tr}_b[ { g}_{ xy}(\boldsymbol{k})  ] )^2
  } $. 
Similarly, for the most commonly studied model of twisted bilayer graphene (TBG) with two occupied bands, it has been shown~\cite{Xie_2020} that 
\begin{equation} \label{eq:bound}
	\frac{1}{4\pi} \int d^2 {\boldsymbol{k}} \mathrm{Tr}_b[g_{xx}(\boldsymbol{k}) + g_{yy}(\boldsymbol{k})] \geq |e_2|,
\end{equation}
where $e_2$ is the Euler number, a topological invariant found in models with $C_{2z}T$ symmetry \cite{Ahn_2019, Guan_2022}. Since $ \int d^2{\boldsymbol{k}} \textrm{Tr}_b[ { g}_{ \mu\nu}(\boldsymbol{k})  ]$ can be determined by the photon correlation measurements, Eq.~\eqref{eq:main_eq}, we can in both cases provide an upper bound to the topological invariant.

{\it Inequalities involving the spin Chern numbers.}
While $ \mathbb{Z}_2$ insulators have zero Chern number \cite{Kane_2005, Bernevig_2006}, it is interesting to ask whether our method can provide information on the spin Chern number, which we define as $ \mathcal{C}_s = \mathcal{C}_\uparrow - \mathcal{C}_\downarrow$. This is meaningful if the model Hamiltonian is of the form $\hat{\mathcal{H}}(\boldsymbol{k}) = \text{diag}({h}_\uparrow( \boldsymbol{k}), {h}_\downarrow (\boldsymbol{k}) )\equiv \text{diag}({h}( \boldsymbol{k}), {h}^* (-\boldsymbol{k}) )$. For such a model with time reversal symmetry (TRS) and inversion symmetry (IS), we can prove the following chain of inequalities 
\begin{align} \label{eq:main_ineq}
  &\pi |\mathcal{C}| \leq \pi ( | { \mathcal{C}}_{ \uparrow} | + |{ \mathcal{C}}_{ \downarrow} | ) = 2\pi |{ \mathcal{C}}_{ \sigma} | \leq  \int_{ }^{ } d^2 {\boldsymbol{k} } \sqrt{ \textrm{det}( \textrm{Tr}_b[ g( \boldsymbol{k} )] )} \nonumber \\ 
  &\quad = 2 \int_{ }^{ } d^2{ \boldsymbol{k}  }  \sqrt{ \textrm{det} (\text{Tr}_\alpha [ g^{ (\alpha\sigma), (\alpha\sigma)}( \boldsymbol{k} )]) }.
\end{align}
TRS and IS imply $ { g}^{ (\alpha\uparrow), (\alpha\uparrow)} ( \boldsymbol{k} ) = { g}^{ (\alpha \downarrow), (\alpha\downarrow)} ( - \boldsymbol{k} )$ and $ { g}^{ (\alpha \sigma), (\alpha\sigma)} ( - \boldsymbol{k} ) = { g}^{ (\alpha\sigma), (\alpha\sigma)} (  \boldsymbol{k} )$, which gives $  { g}^{ (\alpha \uparrow), (\alpha\uparrow)} (  \boldsymbol{k} ) = { g}^{ (\alpha \downarrow), (\alpha\downarrow)} (  \boldsymbol{k} )$. Together with Eq.~\eqref{eq:metric_curvature}, this establishes that $\pi|{ \mathcal{C}}_{ \sigma} | \leq \frac{1}{2} \textrm{vol}_g \leq \frac{1}{2} \textrm{vol}_{\tilde{g}}$ (see SM).

{\it Improved bounds from energy resolution.}
We now explore the potential of energy-resolved measurements, as sketched in Fig.~\ref{fig:setup}(c), for the extraction of the quantum geometry. The energy resolution  allows to insert additional inequalities into Eqs.~\eqref{eq:metric_curvature} and~\eqref{eq:bound}, 
which enables a more complete characterization of the topology of the system. From Eq.~\eqref{eq:main_eq} we can extract the QGT at iso-energy surfaces, $\{ \boldsymbol{k} \in \text{BZ} , \epsilon_{\beta\alpha}(\boldsymbol{k}) - \omega_L =\omega , \alpha\in [1,M] \text{ and } \beta\in [M+1,N]  \}$, corresponding to the contribution $\int d^2{\boldsymbol{k}} \sum^{}_{\alpha, \beta} \delta( \omega -  ( { \epsilon}_{ \beta\alpha}(\boldsymbol{k}) - \omega_L)  ) { f}^{ \mu} {{ f}^{ \nu}}^{ *} \mathcal{A}^{ \alpha\beta}_{ \mu}(\boldsymbol{k}) \mathcal{A}^{ \beta\alpha}_{ \nu} (\boldsymbol{k})$. Employing the Cauchy-Schwarz inequality, we find that
\begin{align}\label{eq:DD_CS}
    &\int_{-\infty}^{\infty} d \omega \bigg[ \Xi_{xx}(\omega) \Xi_{yy}(\omega) - \Big( \text{Re} \Xi_{xy}(\omega) \Big)^2   \bigg]^{1/2} 
    \equiv I_\text{ub} \leq \mathrm{vol}_{\tilde{g}} ,
\end{align}
where $ \Xi_{\mu\nu}(\omega) \equiv \sum_{\alpha=1}^{M} \sum_{\beta=M+1}^N \int d^2 \boldsymbol{k} \delta(\omega - (\epsilon_{\beta\alpha}(\boldsymbol{k}) -\omega_L)  ) \mathcal{A}^{ \alpha\beta}_{\mu }(\boldsymbol{k}) \mathcal{A}^{ \beta\alpha}_{ \nu }(\boldsymbol{k})$. Similarly, we can derive an upper bound to $\pi | \mathcal{C}|$,
\begin{equation}
  \begin{aligned}
    \label{eq:Chern_bound_gen}
    \pi | \mathcal{C}| &\leq \int_{-\infty }^{ \infty}  \: d{ \omega} \Big| \text{Im} \Xi_{xy}(\omega)  \Big| \equiv I_\text{lb}.
  \end{aligned}
\end{equation}
$I_\text{lb}$ is a strict upper bound for $ \pi | \mathcal{C}|$ if $ \text{Im} \Xi_{xy}(\omega)$ evaluates to a negative number on certain iso-energy surfaces. 
Although we have established that 
\begin{equation} \label{eq:ineqs}
\begin{aligned}
	&\pi |\mathcal{C}| \leq I_{\mathrm{lb}} \text{ and } I_{\mathrm{ub}} \leq \mathrm{vol}_{\tilde{g}}, 
\end{aligned}
\end{equation}
we are only able to prove the inequality $ I_{\mathrm{lb}} \leq  \text{vol}_g \leq I_{\mathrm{ub}}$ in the case of a two-band system.
Nevertheless, the knowledge of $I_{\text{lb}}$ and $I_{\text{ub}}$ enables a more precise characterization of geometrical properties. 

\begin{figure*}[htpb]
	\includegraphics[width=\textwidth]{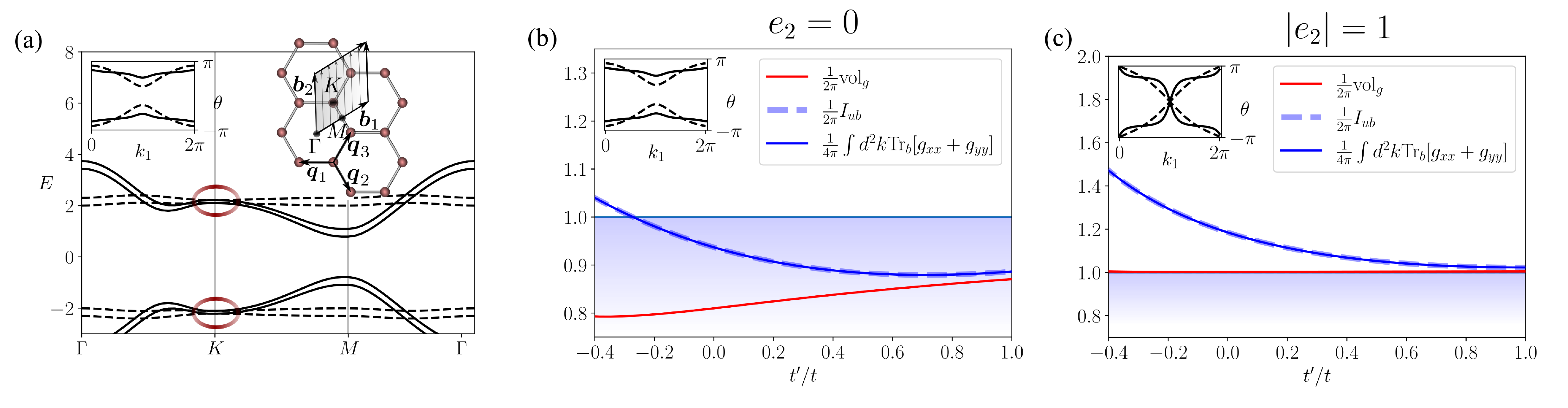}
  \caption{ Geometrical quantities for TBG plotted as a function of $t'/t$. The trivial phase is indicated with a blue shading -- whenever one of the plotted quantities drops below the light-blue line, Eq.~\eqref{eq:bound} implies $e_2=0$. (a) Band structure in the trivial phase where the nodal point at K is gapped due to $\xi =0.8$ (encircled points). (b) Trivial phase ($e_2=0$) with $\xi=0.8$. The inset shows the Wilson loop in the Moiré Brillouin zone for $t'/t=-0.4$ (solid) and $t'/t=1$ (dashed) which are gapped for $\xi=0.8$. (c) Non-trivial phase ($|e_2|=1$) with $\xi=0$ and unit winding of the Wilson loop for all values of $t'/t$.  }
  \label{fig:Euler}
\end{figure*}

{\it Additional information from quantum metric bounds.} 
As a minimal model of a Chern insulator, we consider the QWZ model \cite{Asboth_2016} which corresponds to ${h}(\boldsymbol{k}) = \boldsymbol{d}(\boldsymbol{k}) \cdot \boldsymbol{\sigma}$ with $\boldsymbol{d}(\boldsymbol{k}) = ( \sin(k_x), \sin(k_y), u + \cos( k_x) + \cos( k_y))$ in the space of $s$ and $p$ orbitals. The model displays a non-trivial phase with $| \mathcal{C}|=1$ if the staggered on-site potential $u \in [-2, 2]$, while $ \mathcal{C}=0$ otherwise. A nonzero spin-orbit coupling can be modeled by defining $\hat{\mathcal{H}}(\boldsymbol{k}) = \text{diag}(h(\boldsymbol{k}), h^* (-\boldsymbol{k}) )-\Delta s_y \sigma_y $, where $s_y$ is a Pauli matrix in spin space. We will refer to this model as the SOC model in the following.  

{\it QWZ and SOC models.} 
Figure~\ref{fig:inequalities}(a-b) shows results for the model without and with spin-orbit coupling, respectively. As is evident from Fig.~\ref{fig:inequalities}(a), the second inequality in Eq.~\eqref{eq:ineqs} is saturated for the QWZ model while, on the other hand, $I_\text{lb} = \text{vol}_g$ holds at most values of $u$. This is related to the metric-curvature correspondence noted in Ref.~\cite{Ozawa_2021}, which for the case of two bands is $\sqrt{\text{det}(g)} = | \frac{\mathcal{F}_{xy}}{2} |$. The fact that $I_\text{lb}=\text{vol}_g$ does not hold at all values of $u$ can be attributed to sign changes of $\mathcal{F}_{xy}$ along iso-energy surfaces (see SM).

As shown in Fig.~\ref{fig:inequalities}(b), the introduction of a small SOC $\Delta$ leads to a more pronounced peak structure of $ \text{vol}_{\tilde{g}}$ near $u=2$.  
The band touching at $(\pi, \pi)$ turns into a band touching at several momentum points once the spin degeneracy of the model is lifted, which subsequently contributes to a larger QGT. Nevertheless, we still find that $ \tfrac{1}{2\pi}\mathrm{vol}_{\tilde{g}}$ provides a useful upper bound of $\mathcal{C}_\sigma$ away from the transition.

{\it Three band model.} 
Next we analyze the three band model from Ref.~\cite{Ozawa_2021} with $\hat{ \mathcal{H}}_{11} ( \boldsymbol{k} )=-2t_{dd} (\cos(k_x) + \cos(k_y)) + \delta$, $\hat{ \mathcal{H}}_{12} ( \boldsymbol{k} )=\hat{ \mathcal{H}}_{21} ( \boldsymbol{k} )^*=2i t_{pd} \sin(k_x)$, $\hat{ \mathcal{H}}_{13} ( \boldsymbol{k} )=\hat{ \mathcal{H}}_{31} ( \boldsymbol{k} )^*=2i t_{pd} \sin(k_y)$, $\hat{ \mathcal{H}}_{22} ( \boldsymbol{k} ) = 2 t_{pp} \cos(k_x) - 2 t_{pp}^{'} \cos(k_y) $, $\hat{ \mathcal{H}}_{23} ( \boldsymbol{k} ) = \hat{ \mathcal{H}}_{32} ( \boldsymbol{k} )^*= i \Delta$, $\hat{ \mathcal{H}}_{33} ( \boldsymbol{k} ) = 2 t_{pp}\cos(k_y) - 2 t_{pp}^{'} \cos(k_x) $, $t_{dd} = t_{pd} = t_{pp}=1$ and $ \delta = -4 t_{dd} + 2 t_{pp} + \Delta - {2 t_{pp} \Delta /(4 t_{pp} + \Delta})$, $ t_{pp}^{'} = t_{pp}\Delta/( 4 t_{pp} + \Delta )$ and compute the QGT with respect to the lowest band. Fig.~\ref{fig:inequalities}(c) shows that -- similarly to the QWZ model --  $I_\text{lb}$ closely follows $ \mathrm{vol}_g$ over the entire trivial phase, and to some extent also in the non-trivial phase. Since $ \sqrt{\text{det}(g)}$ can differ from $| \mathcal{F}_{xy}|/2$ in models with more than two bands, $I_\text{lb}$ is a less good estimate of $ \text{vol}_{g}$. Still, $I_\text{lb}$ can be used to estimate the location of the topological transition with high accuracy. 

{\it Twisted bilayer graphene (TBG).} 
As a further application, we consider the 4-band tight-binding model of TBG from Ref.~\cite{Xie_2020},
\begin{equation}
\begin{aligned}
  \label{eq:ham}
  \hat{\mathcal{H}}( \boldsymbol{k} ) &=  { \mu}_{ z} ( \Delta { \sigma}_{ 0} +\xi { \sigma}_{ z }) + \mu_0 \boldsymbol{\rho}(\boldsymbol{k}) \cdot \boldsymbol{\sigma} 
   - 2\lambda \mu_y \sigma_z f(\boldsymbol{k}), 
\end{aligned}
\end{equation}
where $ { \mu}_{(0), x,y,z} $ and $ {\sigma}_{(0), x,y,z} $ are (identity matrices) Pauli matrices acting in the space of orbital and sublattice degrees of freedom, respectively. $\rho_{1,2}(\boldsymbol{k}) = \sum^{3}_{i=1} [ t \cos / \sin( { \boldsymbol{\delta} }_{ i} \cdot \boldsymbol{k} ) + t' \cos / \sin( -2 { \boldsymbol{\delta} }_{ i} \cdot \boldsymbol{k} )] $, $\rho_3(\boldsymbol{k})=0$ and $f(\boldsymbol{k}) = \sum_{i=1}^3 \sin( \boldsymbol{d}_i \cdot \boldsymbol{k})$. With $\boldsymbol{a}_{1,2}$ representing the real-space Moiré lattice unit vectors, we have the nearest neighbor vectors $ { \boldsymbol{\delta} }_{ 1} = \frac{ 1}{ 3} { \boldsymbol{a} }_{ 1} + \frac{ 2}{ 3} { \boldsymbol{a} }_{ 2} $, $ { \boldsymbol{\delta} }_{ 2} = - \frac{ 2}{ 3} { \boldsymbol{a} }_{ 1} - \frac{ 1}{ 3} { \boldsymbol{a} }_{ 2}  $, $ { \boldsymbol{\delta} }_{ 3} =  \frac{ 1}{ 3} { \boldsymbol{a} }_{ 1} + \frac{ 2}{ 3} { \boldsymbol{a} }_{ 2}  $ and the second nearest neighbor vectors $ { \boldsymbol{d} }_{ 1}=\boldsymbol{a}_1 $, $ { \boldsymbol{d} }_{ 2}=\boldsymbol{a}_2 $, $ { \boldsymbol{d} }_{ 3}=-\boldsymbol{a}_1-\boldsymbol{a}_2$. In order to make the bands as flat as possible, Xie {\it et al.} chose $ t' = - \frac{ t}{ 3} $, $ \lambda = (2/ \sqrt{ 27} )t$ and $\Delta = 0.15 t$ with $t=1$ \cite{Xie_2020}. A non-zero $\xi$ opens a gap between the otherwise degenerate occupied and unoccupied bands at $K$, rendering the topological phase trivial (see Fig.~\ref{fig:Euler} (a)). 

The nontrivial topology in this model manifests itself in Wilson loop winding which is seen as a crossing in the loop diagrams shown in the insets of Fig.~\ref{fig:Euler} \cite{Song_2019, Ahn_2019}. In the same figure, we present geometrical quantities for the two phases of the model. In view of Eq.~\eqref{eq:bound}, Fig.~\ref{fig:Euler}(b) shows that for most values of $t'/t$, we can determine whether the system is in a topologically trivial state by looking at momentum integrals over $ g_{ii}$ $(i=x,y)$.

In the non-trivial case depicted in Fig.~\ref{fig:Euler}(c), we see the same trend of $ { g}_{ ii} $ increasing with increasing bandwidth, but find a remarkable agreement between $ \frac{ 1}{ 4 \pi} \int_{ }^{ }  d^2 \boldsymbol{k} { \textrm{Tr}}_{ b} [ { g}_{ xx} + { g}_{ yy} ]$ and $ | { e}_{ 2} |$ at smaller bandwidths, which is consistent with Ref.~\cite{Ozawa_2018}. We also observe that $\tfrac{1}{2\pi} { \textrm{vol}}_{ g} $ matches $ |{ e}_{ 2} |$ almost perfectly for all bandwidths in the non-trivial phase.  Although the relationship between $ \tfrac{1}{2\pi}{ \textrm{vol}}_{ g} $ and $ | { e}_{ 2} |$ is unclear, we numerically demonstrate that $\tfrac{1}{2\pi}{\textrm{vol}}_{g}$ provides an upper bound to $|e_2|$ for all $t/t'$.

{\it Conclusions.} 
We have devised a way of extracting quantum geometry, and indirectly topology, by means of 
a cavity QED set-up combined with a heterodyne detection scheme and 
the localization dichotomy. The power of this scheme was demonstrated with applications to paradigmatic models hosting interesting geometry and topology. Utilizing the energy resolution of the method we provided improved markers of the system's geometry. A future prospect is the application of our scheme to 3D materials (thin films) with nontrivial geometrical properties. While there is no Chern number in 3D, the delocalization of Wannier orbitals can be detected. Furthermore, in systems with Berry curvature dipoles and zero Chern number, the metric-curvature correspondence still enables a characterization of the Berry curvature dipole strength \cite{Xu_2018, Ye_2023}. 

{\it Acknowledgments}
This work has been supported by the European Research Council through ERC Consolidator Grant No. 724103.
The calculations have been performed on the Beo05 cluster at the University of Fribourg. M.S. thanks the Swiss National Science Foundation SNSF for its support with an Ambizione grant (project no. PZ00P2-193527).

\bibliography{bibliography.bib}

\end{document}

\title{
Quantum optics measurement scheme for quantum geometry \\ and topological invariants -- Supplementary material
}
\maketitle

\section{Outline}

In this supplementary material, we review some properties of quantum geometry which are relevant for the theory in the main manuscript. In particular, in Sec.~\hyperref[sec:Wilson]{II.B}, the Wilson loop approach to computing topological invariants is formulated in a way that applies to the Euler number. In Sec.~\hyperref[sec:DD_bounds]{II.C}, we prove some statements given in the main manuscript regarding inequalities for the QWZ model. We also present a detailed derivation of a central result of the paper, Eq.~(3), in Secs.~\hyperref[sec:QO]{III-V} following Vogel and Welsch \cite{Vogel_2006}.

\section{Geometry} \label{sec:topology}

\subsection{Quantum geometric tensor and quantum metric}
The definition of the quantum geometric tensor (QGT) has been given in Eq.~(4) of the manuscript. Here, we provide an alternative expression which is more amenable to numerical treatment. Considering the eigenvalue equation for the periodic part of the Bloch state, $\mathcal{\hat{H}}(\boldsymbol{k}) | u_{\boldsymbol{k}, \alpha} \rangle = \epsilon_{\alpha}(\boldsymbol{k}) | u_{\boldsymbol{k},\alpha} \rangle$, and computing the $k$-derivative of the Bloch states by perturbation theory, we can show that
\begin{equation} \label{eq:vel_to_Berry}
	\langle u_{\boldsymbol{k},\beta} | \partial_\mu | u_{\boldsymbol{k},\alpha}  \rangle = \frac{ \langle u_{\boldsymbol{k},\beta} | \partial_\mu\mathcal{H}(\bs{k}) | u_{\boldsymbol{k},\alpha} \rangle}{\epsilon_\alpha(\bs{k}) - \epsilon_\beta(\bs{k})} = -i { \mathcal{A}}^{ \beta\alpha}_{ \mu} ( \boldsymbol{k} ),
\end{equation}
which is valid for $\epsilon_{\alpha}\neq \epsilon_{\beta}$. Hence, from Eq.~(4) in the manuscript, we may write
\begin{equation} \label{eq:QGT_GI}
	Q_{\mu\nu}^{\alpha\beta}(\bs{k}) = \sum_{\gamma=M+1}^N \frac{ \langle u_{\boldsymbol{k},\alpha} | \partial_\mu \mathcal{H}(\bs{k}) | u_{\boldsymbol{k},\gamma } \rangle \langle u_{\boldsymbol{k},\gamma} | \partial_\nu \mathcal{H}(\bs{k}) | u_{\boldsymbol{k},\beta}  \rangle }{(\epsilon_\alpha(\bs{k}) - \epsilon_{\gamma}(\bs{k})  )(\epsilon_\beta(\bs{k}) - \epsilon_\gamma (\bs{k})  )},
\end{equation}
where $\alpha,\beta \in [1,M]$ run over occupied bands. The real and imaginary parts of ${Q}_{ \mu \nu}(\boldsymbol{k})$ are related to the quantum metric ($g^{\alpha\beta}_{\mu\nu}(\boldsymbol{k})$) and Berry curvature ($\mathcal{F}^{\alpha\beta}_{\mu\nu}(\boldsymbol{k})$) by
$Q_{\mu\nu}^{\alpha\beta}( \boldsymbol{k}) = g_{\mu\nu}^{\alpha\beta}(\boldsymbol{k}) - \frac{i}{2}\mathcal{F}_{\mu\nu}^{\alpha\beta}(\boldsymbol{k})$.
 These quantities 
 can also be defined as the symmetric and antisymmetric parts of $ { Q}_{ \mu \nu}(\bs{k}) $ \cite{Ding_2022},
\begin{equation}\label{eq:symmetric_ind}
  \begin{aligned}
    g_{\mu\nu}(\boldsymbol{k}) &= \frac{ 1}{ 2 } [ {Q}_{ \mu \nu}(\boldsymbol{k}) + { Q}_{ \mu \nu}^\dagger (\boldsymbol{k})  ] , \\
    { \mathcal{F}}_{ \mu \nu}(\boldsymbol{k}) &= i [ {Q}_{ \mu \nu}(\boldsymbol{k}) - { Q}_{ \mu \nu}^\dagger (\boldsymbol{k})  ] .
  \end{aligned}
\end{equation}
Since $Q_{\mu\nu}^{ab,*}(\boldsymbol{k}) = Q_{\nu\mu}^{ba}(\boldsymbol{k})$ we find that $g_{\mu\nu}^{ab}(\bs{k})$ ($\mathcal{F}_{\mu\nu}^{ab}(\bs{k}) $) is symmetric (anti-symmetric) in $\mu$ and $\nu$. When $\alpha=\beta$, Eq.~\eqref{eq:QGT_GI} does not depend on the gauge of the Bloch functions -- i.e., it is invariant under $| u_{\boldsymbol{k}, \alpha} \rangle \rightarrow e^{i\phi_\alpha (\boldsymbol{k})} | u_{\boldsymbol{k}, \alpha} \rangle $ --  and we will therefore use this formula in most numerical calculations. For the computation of the Euler number, we would however need $\alpha\neq \beta$ \cite{Ahn_2019}.

\begin{figure}[t]
  \centering
  \subfigure{\includegraphics[width=0.45\linewidth]{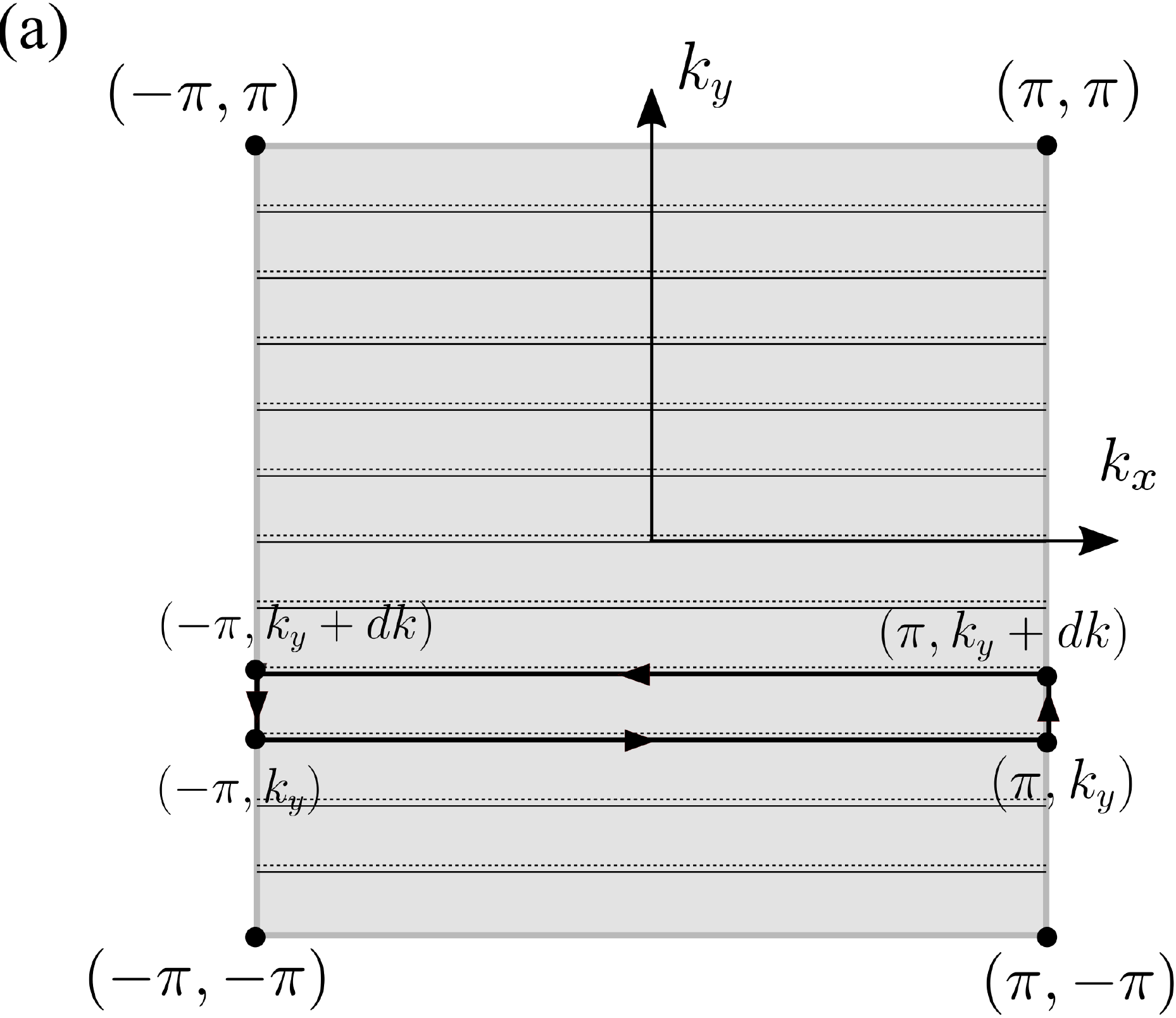}} \hfill
    \subfigure{\includegraphics[width=0.45\linewidth]{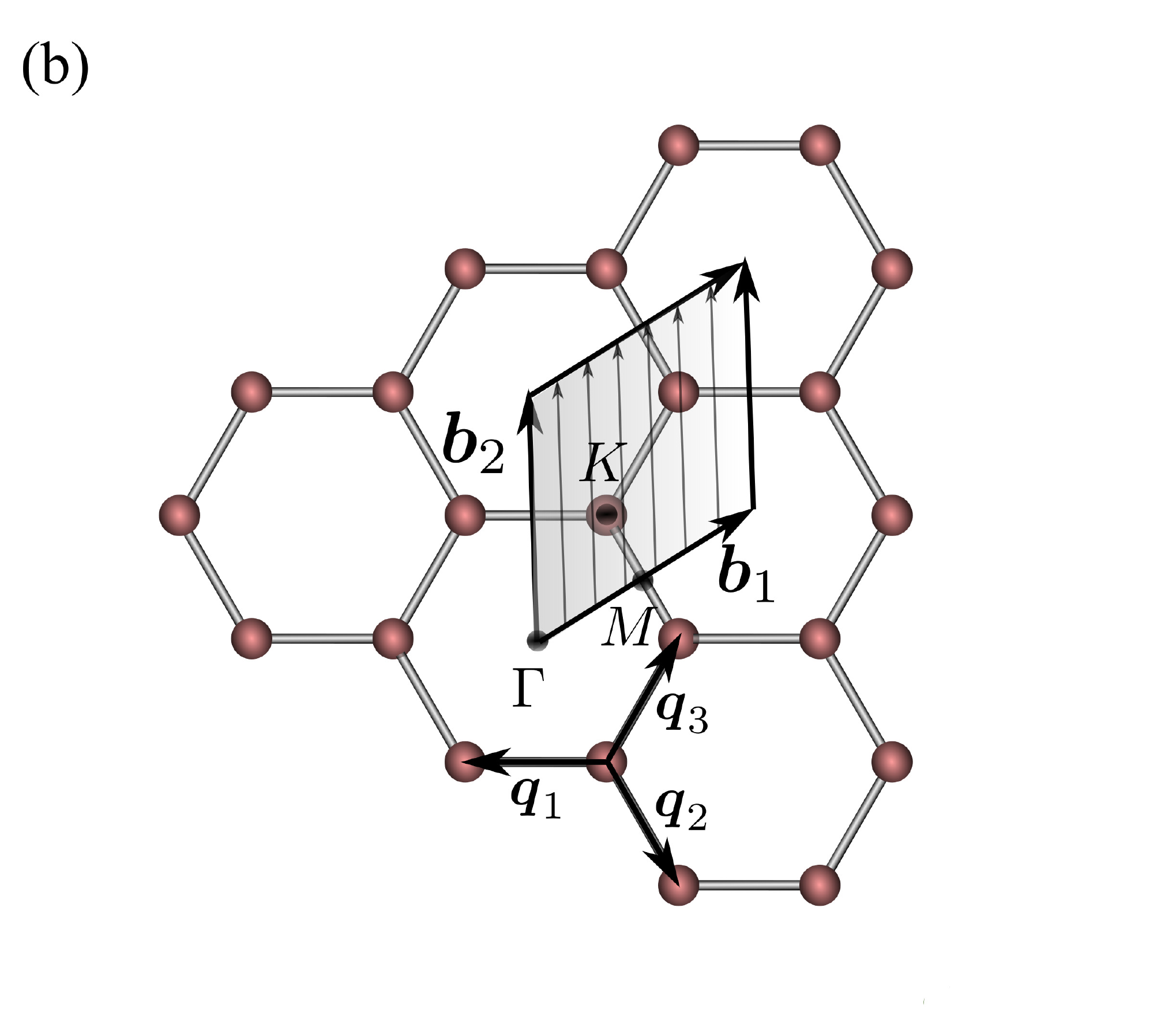}} 
  \caption{(a) Illustration showing how the BZ can be split into rectangles consisting of large Wilson loops in which we can choose a smooth gauge for the Bloch states. 
  (b) In the Moiré BZ of the TBG model, the Wilson lines are indicated by vertical lines.}
  \label{fig:images_wilson_drawing}
\end{figure}

\subsection{Computing the Euler number from the Wilson loop} \label{sec:Wilson}

In order to compute the Euler number, we draw on some previous results presented in Refs.~\cite{Bouhon_2020,Asboth_2016}. 
Using the non-Abelian Berry connection defined in the manuscript, we define the Wilson loop operator as 
\begin{equation}
  \hat{\mathcal{W}} = \hat{\mathcal{P}} \text{exp}\Big\{ -i\oint_{\mathcal{C}} d \boldsymbol{q} \cdot \overrightarrow{\mathcal{A}}( \boldsymbol{q} ) \Big\},
\end{equation}
where $\hat{\mathcal{P}}$ is the path-ordering operator and $\mathcal{C}$ denotes a closed loop in the Brillouin zone (BZ). $\overrightarrow{\mathcal{A}}(\boldsymbol{k}) \equiv ( \mathcal{A}_1 (\boldsymbol{k}), \mathcal{A}_2 (\boldsymbol{k}))$ where $\mathcal{A}_\mu (\boldsymbol{k})$ is a matrix in the space of \emph{occupied} states. 
From the non-Abelian Berry curvature 
\begin{equation}
	\mathcal{F}_{\mu\nu}^{\alpha\beta}(\boldsymbol{k})= \partial_{\mu} \mathcal{A}_{\nu}^{\alpha\beta}(\boldsymbol{k}) - \partial_{\nu} \mathcal{A}_{\mu}^{\alpha\beta}(\boldsymbol{k}) - i [\mathcal{A}_{\mu}(\boldsymbol{k}), \mathcal{A}_{\nu}(\boldsymbol{k}) ]^{\alpha\beta}
\end{equation}
we define the ``Euler curvature" as  
\begin{equation}
	\text{Pf}(\mathcal{F}_{xy}(\boldsymbol{k})) = \partial_x \text{Pf}( \mathcal{A}_y (\boldsymbol{k})) - \partial_y \text{Pf}( \mathcal{A}_x (\boldsymbol{k})),
\end{equation}
where we have used the linearity of the Pfaffian for skew-symmetric $2\times 2$ matrices and that the Pfaffian of a commutator of such matrices is zero. These properties follow from Refs.~\cite{Ahn_2019, Xie_2020}, who showed how it is possible to find a gauge for which $\mathcal{A}_\mu (\boldsymbol{k})$ is skew-symmetric. We may then apply Stokes theorem
\begin{equation} \label{eq:Stokes}
	\int \int_S  d^2 \boldsymbol{k} \text{Pf}(\mathcal{F}_{xy}(\boldsymbol{k})) = \oint_{\partial S} d \boldsymbol{q} \cdot \text{Pf}(\overrightarrow{\mathcal{A}}(\boldsymbol{q})) ,
\end{equation}
where we have defined $\text{Pf}(\overrightarrow{\mathcal{A}}(\boldsymbol{k})) \equiv ( \text{Pf}(\mathcal{A}_1 (\boldsymbol{k})), \text{Pf}( \mathcal{A}_2 (\boldsymbol{k})))$. This equation is valid for regions of the BZ where we can define a consistent gauge. For the following construction, we restrict ourselves to the BZ of the square lattice with lattice parameter $a=1$, as shown in Fig.~\ref{fig:images_wilson_drawing}, and split the BZ into rectangles with width $ 2 \pi$ in the $k_x$ direction and height $ dk$ in the $k_y$ direction -- the extension to arbitrary 2D lattices is straightforward. We define a large Wilson loop across the entire BZ for $k_y$ fixed to be $\mathcal{W}_{k_y} = \text{exp} \{ -i \oint_{k_y} d \boldsymbol{q} \cdot \overrightarrow{\mathcal{A}}( \boldsymbol{q} )  \}$.
Here, the path corresponds to the thin rectangle with lower edge at position $k_y$ (Fig.~\ref{fig:images_wilson_drawing}(a)). Since $ \oint_{k_y} d \boldsymbol{q} \cdot \overrightarrow{\mathcal{A}}( \boldsymbol{q} ) $ is skew-symmetric, we can parametrize it as
\begin{equation}
  \oint_{k_y} d \boldsymbol{q} \cdot \overrightarrow{\mathcal{A}}( \boldsymbol{q} )  = -\alpha( { k}_{ y} ) { \sigma}_{ y},
\end{equation}
and we may therefore write $\mathcal{W}_{k_y} = \text{exp} \{ {i \alpha(k_y) \sigma_y} \} $, which means that $\alpha(k_y) = \text{Pf} [\text{log} \mathcal{W}_{k_y}]$. Now, for a single rectangle in Fig~\ref{fig:images_wilson_drawing}(a) and by Eq.~\eqref{eq:Stokes}, we get
\begin{equation}
	i[ \alpha(k_y) - \alpha(k_y + dk)] = \oint_{k_y} d\boldsymbol{q} \cdot \text{Pf}( \overrightarrow{\mathcal{A}}(\boldsymbol{k}) ) - \oint_{k_y + dk} d\boldsymbol{q} \cdot \text{Pf}( \overrightarrow{\mathcal{A}}(\boldsymbol{k}) ) = \iint_{\square_{k_y}} d^2 \boldsymbol{k} \text{Pf}(\mathcal{F}_{xy}(\boldsymbol{k})) ,
\end{equation}
where $\square_{k_y}$ is the area of the BZ between $k_y$ and $k_y + dk$. If we sum up all of the rectangles in the BZ, we find that 
\begin{equation}
	i[ \alpha(-\pi) - \alpha(\pi)] =  \iint_{ \text{BZ} } d^2 \boldsymbol{k} \text{Pf}(\mathcal{F}_{xy}(\boldsymbol{k})).
\end{equation}
Since the Wilson operator, $\mathcal{W}_{k_y}$, must coincide at $ { k}_{ y} = \pi$ and $ k_y= - \pi$, we have that $ \alpha( \pi) = 2\pi e_2 + \alpha(-\pi)$ with $e_2$ an integer. Therefore, 
\begin{equation}
	e_2 =  \frac{i}{2\pi} \iint_{ \text{BZ} } d^2 \boldsymbol{k} \text{Pf}(\mathcal{F}_{xy}(\boldsymbol{k})).
\end{equation}

The numerical procedure for calculating the topological invariant by means of the Wilson loop method is detailed in Ref.~\cite{Rui_2011}. For the TBG case, we traverse all $k$ points along the $\boldsymbol{b}_1$ axis and compute the eigenvalues of the large Wilson loops along the $\boldsymbol{b}_2$ axis and subsequently obtain the winding number, which is $|e_2|$. This is indicated in Fig.~\ref{fig:images_wilson_drawing}(b), where we show the reciprocal lattice of TBG overlaid with its irreducible BZ and large Wilson loops (vertical lines).

\subsection{Bounds using energy resolution} \label{sec:DD_bounds}

We want to prove the assertion that 
\begin{equation}
	\pi |\mathcal{C}| \leq I_\text{lb} \leq \text{vol}_g \leq I_\text{ub} \leq \text{vol}_{\tilde{g}}
 \end{equation}
holds for a two-band system. Having established $\pi |\mathcal{C}| \leq I_\text{lb}$ and $I_\text{ub} \leq \text{vol}_{\tilde{g}}$ in the manuscript, we must prove $ \text{vol}_g \leq I_{\text{ub}}$ and $I_{\text{lb}} \leq \text{vol}_{g}$. Note that in systems with more than two bands, $\Xi_{\mu\nu} (\omega) \equiv \sum_{\alpha=1}^{M} \sum_{\beta=M+1}^N \int d^2 \boldsymbol{k} \delta(\omega - (\epsilon_{\beta\alpha}(\boldsymbol{k}) -\omega_L)  ) \mathcal{A}^{ \alpha\beta}_{\mu }(\boldsymbol{k}) \mathcal{A}^{ \beta\alpha}_{ \nu }(\boldsymbol{k})$ may pick up contributions from several pairs of bands with energy separation $\omega$. However, in the two band case we may decompose the full $k$ space integral as $ \int_{ }^{ }  \: d^2{ \boldsymbol{k}} = \int d\omega \int d^2 \boldsymbol{k} \delta(\omega - (\epsilon_{21}(\boldsymbol{k}) -\omega_L ))$. Without loss of generality, let us set $\omega_L=0$ in the following. We may then apply the Cauchy-Schwarz inequality and obtain

\begin{equation}\label{eq:series_of_ineqs}
\begin{aligned}
	\text{vol}_g &= \int_{-\infty}^{\infty} d\omega \int d^2 \boldsymbol{k} \delta(\omega - \epsilon_{21}(\boldsymbol{k})) \big[ g_{xx}(\boldsymbol{k}) g_{yy}(\boldsymbol{k}) - g_{xy}^2(\boldsymbol{k})  \big]^{1/2} \\
	&=\int_{-\infty}^{\infty} d\omega \int d^2 \boldsymbol{k} \delta(\omega - \epsilon_{21}(\boldsymbol{k})) \Big[ \Big(\sqrt{g_{xx}(\boldsymbol{k}) g_{yy}(\boldsymbol{k})} - g_{xy}(\boldsymbol{k}) \Big) \cdot  
	\Big(\sqrt{g_{xx}(\boldsymbol{k}) g_{yy}(\boldsymbol{k})} + g_{xy}(\boldsymbol{k}) \Big) \Big]^{1/2} \\
	&\leq \int_{-\infty}^{\infty} d\omega  \Big[ \int d^2 \boldsymbol{k} \delta(\omega - \epsilon_{21}(\boldsymbol{k})) \Big(\sqrt{g_{xx}(\boldsymbol{k}) g_{yy}(\boldsymbol{k})} - g_{xy}(\boldsymbol{k}) \Big) \cdot 
	\int d^2 \boldsymbol{k} \delta(\omega - \epsilon_{21}(\boldsymbol{k})) \Big(\sqrt{g_{xx}(\boldsymbol{k}) g_{yy}(\boldsymbol{k})} + g_{xy}(\boldsymbol{k})\Big) \Big]^{1/2} \\
	&\leq \int_{-\infty}^{\infty} d\omega \Big[ \int d^2 \boldsymbol{k} \delta(\omega - \epsilon_{21}(\boldsymbol{k})) g_{xx}(\boldsymbol{k}) \int d^2 \boldsymbol{k} \delta(\omega - \epsilon_{21}(\boldsymbol{k})) g_{yy}(\boldsymbol{k})- \Big(\int d^2 \boldsymbol{k} \delta(\omega - \epsilon_{21}(\boldsymbol{k})) g_{xy}(\boldsymbol{k})\Big)^2  \Big]^{1/2}  = I_{\text{ub}}.
\end{aligned}
\end{equation}
In the last line, we have applied the Cauchy-Schwarz inequality once more in writing $ \int d^2 \boldsymbol{k} \delta(\omega - \epsilon_{21}(\boldsymbol{k})) \sqrt{ g_{xx}(\boldsymbol{k}) g_{yy}(\boldsymbol{k}) } \leq \sqrt{  \int d^2 \boldsymbol{k} \delta(\omega - \epsilon_{21}(\boldsymbol{k})) g_{xx}(\boldsymbol{k}) \int d^2 \boldsymbol{k} \delta(\omega - \epsilon_{21}(\boldsymbol{k})) g_{yy}(\boldsymbol{k}) } $. 

Next, let us comment on why $ I_{\text{lb}} < \text{vol}_g$ in some cases, which is related to Eq.~(9) in the manuscript. For a two band model, such as the QWZ model, we have 
\begin{equation} \label{eq:Berry_bound}
\begin{aligned}
	I_{\text{lb}}&=\int_{-\infty}^\infty d\omega |  \text{Im} \Xi_{xy} (\omega)| =  \int_{-\infty}^\infty d\omega \int d^2 \boldsymbol{k} \Big|\delta(\omega - \epsilon_{21}(\boldsymbol{k})) \frac{\mathcal{F}_{xy}(\boldsymbol{k})}{2} \Big| \\
	&\leq \int_{-\infty}^\infty d\omega \int d^2 \boldsymbol{k} \delta(\omega - \epsilon_{21}(\boldsymbol{k}))\Big|\frac{\mathcal{F}_{xy}(\boldsymbol{k})}{2} \Big| = \text{vol}_g.
\end{aligned}
\end{equation}
The last inequality is strict whenever $\mathcal{F}_{xy}(\boldsymbol{k})$ changes sign along one of the fixed energy difference contours, and we will show that this does in fact occur for the transition at $u=0$, but not for the one at $u=2$.  
By noting that band touchings are sources of Berry curvature, we identify the band-touching points at the two topological transitions of interest and assume that there are fixed energy contours in the vicinity of these points with approximately circular paths. It was shown in Ref.~\cite{Bernevig_2013} that for $\hat{h}(\boldsymbol{k}) = \boldsymbol{d}(\boldsymbol{k}) \cdot \boldsymbol{\sigma} $
\begin{equation}
	\mathcal{F}_{xy}(\boldsymbol{k}) = \frac{1}{2|\boldsymbol{d}(\boldsymbol{k})|^3}\epsilon_{abc} d_a(\boldsymbol{k}) \partial_x d_b(\boldsymbol{k}) \partial_y d_c(\boldsymbol{k})=\frac{1}{2|\boldsymbol{d}(\boldsymbol{k})|^3} \tilde{\mathcal{F}}_{xy}(\boldsymbol{k}),
\end{equation}
where 
\begin{equation}
	\tilde{\mathcal{F}}_{xy}(\boldsymbol{k}) = \cos(k_x) \sin^2 (k_y) + \cos(k_y) \sin^2 (k_x) + ( u + \cos(k_x) + \cos(k_y) )\cos(k_x)\cos(k_y)
\end{equation}
for the QWZ model. We know that at $u=0$, the band-touching occurs at $(\pm\pi,0)$ and $(0,\pm\pi)$. Taylor expanding around $(\pi,0)$, we find that 
\begin{equation}
	\tilde{\mathcal{F}}_{xy}(\boldsymbol{k}) \approx \frac{1}{2}(1-u)[ (\pi - k_x)^2 - k_y^2 ],
\end{equation}
which implies that $\mathcal{F}_{xy}(\boldsymbol{k})$ changes sign around this particular contour (similar expressions can be found at $(-\pi,0)$ and $(0,\pm\pi)$). As for the relationship between $I_{\text{lb}}$ and $\text{vol}_g$, let us note that at $u\rightarrow 0$, $d\rightarrow 0$ and thus the main contribution to $I_{\text{lb}}$ comes from the contours close to $(\pm\pi,0)$ and $(0,\pm\pi)$ -- this is what yields $I_{\text{lb}} < \text{vol}_g$, as seen for low $u$ in Fig.~2(a) and (c) of the manuscript.

At $u=2$, we have band-touching at a single point, namely $(\pi, \pi)$. By a Taylor expansion around this point, we find that in this case 
\begin{equation}
	\tilde{\mathcal{F}}_{xy}(\boldsymbol{k}) \approx - \frac{1}{2} u [ (k_x - \pi)^2 + (k_y-\pi)^2 ]
\end{equation}
and hence $\mathcal{F}_{xy}(\boldsymbol{k})$ does not change sign along fixed energy contours, which by Eq.~\eqref{eq:Berry_bound} is consistent with the observations in Fig.~2(a) and (c) of the manuscript. Although we have not demonstrated this behavior for the three band model, the same arguments likely also apply to the plot of $I_{\text{lb}}$ in Fig.~2(c).

\subsection{Inequalities involving the spin Chern numbers}
If we assume time reversal symmetry (TRS) and inversion symmetry (IS), we can prove the following chain of inequalities, 
\begin{equation} \label{eq:main_ineq}
  \pi | \mathcal{C}| \leq \pi ( | { \mathcal{C}}_{ \sigma} | + | { \mathcal{C}}_{ \overline{\sigma}} | ) = 2\pi | { \mathcal{C}}_{ \sigma} |
  \leq  2 \int_{ }^{ } d^2{ \boldsymbol{k}  }  \sqrt{ \textrm{det} (\text{Tr}_\alpha [ g^{ (\alpha\sigma), (\alpha\sigma)}( \boldsymbol{k} )]) }
  =\int_{ }^{ } d^2 {\boldsymbol{k} } \sqrt{ \textrm{det}( \textrm{Tr}_b[ g( \boldsymbol{k} )] )}   ,
\end{equation}
which is Eq.~(7) in the manuscript. By virtue of spin conservation in the model we may label the states by $(\alpha,\sigma)$, where $\sigma$ is the spin quantum number. ($\textrm{Tr}_b[g(\bs{k}) ] = \text{Tr}_{\alpha\sigma} [ g^{ (\alpha\sigma), (\alpha\sigma)}( \boldsymbol{k} )] $.) To study symmetry properties of the quantum metric and the Berry curvature for systems with spin, we follow Ref.~\cite{Sergienko_2004} and find that under TRS, the periodic parts of the Bloch wavefunctions satisfy $|{ u }_{ \boldsymbol{k} , \alpha,\downarrow}^{ *} \rangle = -t( \boldsymbol{k} ) |{ u }_{ - \boldsymbol{k} , \alpha,\uparrow} \rangle $ and $|{ u }_{ \boldsymbol{k} , \alpha,\uparrow}^{ *} \rangle = t( \boldsymbol{k} ) |{ u }_{ - \boldsymbol{k} , \alpha,\downarrow} \rangle $, where $t(\boldsymbol{k})$ is a phase factor. Under inversion symmetry, we find that $ |u_{-\bs{k}, \alpha,\sigma}\rangle = e^{i \phi_{\alpha}(\bs{k})} | u_{\bs{k}, \alpha,\sigma} \rangle$. TRS and IS applied to Eq.~\eqref{eq:symmetric_ind} thus give $ { g}^{(\alpha \uparrow), (\alpha\uparrow)} ( \boldsymbol{k} ) = { g}^{ (\alpha\downarrow), (\alpha\downarrow)} ( - \boldsymbol{k} )$ and $ { g}^{ (\alpha\sigma), (\alpha\sigma)} ( - \boldsymbol{k} ) = { g}^{ (\alpha\sigma), (\alpha\sigma)} (  \boldsymbol{k} )$, respectively.  

To prove the second inequality, we follow the same steps as in Appendix A of Ref.~\cite{Ozawa_2021}, replacing $P(\boldsymbol{k})$ therein by
\begin{equation}
  { P}_{ \sigma}(\boldsymbol{k}) = \sum^{M}_{\alpha=1} | { u }_{\boldsymbol{k}, \alpha,\sigma} \rangle \langle { u }_{ \boldsymbol{k},\alpha, \sigma}  | ,
\end{equation} 
where $M$ is the number of occupied bands of one spin species. By defining $ | v_{\bs{k}, \sigma} \rangle = ( | v_{\bs{k},1,\sigma} \rangle, | v_{\bs{k},2,\sigma} \rangle, ..., | v_{\bs{k},M,\sigma} \rangle)^T $ and $ | w_{\bs{k}, \sigma} \rangle = ( | w_{\bs{k},1,\sigma} \rangle, | w_{\bs{k},2,\sigma} \rangle, ..., | w_{\bs{k},M,\sigma} \rangle)^T $ where
\begin{equation}
\begin{aligned}
	| v_{\bs{k},\alpha,\sigma} \rangle &= (1- P_\sigma (\bs{k})) | \partial_\mu u_{\bs{k}, \alpha, \sigma} \rangle \\
	| w_{\bs{k},\alpha,\sigma} \rangle &= (1- P_\sigma (\bs{k}) ) | \partial_\nu u_{\bs{k}, \alpha, \sigma} \rangle \\ 
\end{aligned}
\end{equation}
and applying the Cauchy-Schwarz inequality, 
\begin{equation}
	\langle v_{\bs{k}, \sigma} | v_{\bs{k}, \sigma}  \rangle \langle w_{\bs{k}, \sigma} | w_{\bs{k}, \sigma} \rangle \geq |\langle v_{\bs{k}, \sigma} | w_{\bs{k}, \sigma} \rangle |^2
\end{equation}
we may show that 
\begin{equation} \label{eq:detg_spin}
	\sqrt{\text{det}( \text{Tr}_\alpha [ g^{(\alpha\sigma), (\alpha\sigma)}(\bs{k}) ] ) } \geq \frac{1}{2}  \text{Tr}_\alpha [ \mathcal{F}_{xy}^{(\alpha\sigma), (\alpha\sigma)}(\bs{k})],
\end{equation}
where we have used Eq.~\eqref{eq:QGT_GI} and Eq.~\eqref{eq:symmetric_ind}. As the spin-Chern number is explicitly defined as  
\begin{equation}
	\mathcal{C}_\sigma = \frac{1}{2\pi} \int d^2 \bs{k}  \text{Tr}_\alpha [ \mathcal{F}_{xy}^{(\alpha\sigma)(\alpha\sigma)}(\bs{k})]
\end{equation}
we see that upon performing the $k$-integral on both sides of Eq.~\eqref{eq:detg_spin} that
\begin{equation}
	\pi | { \mathcal{C}}_{ \sigma} |
  \leq \int_{ }^{ } d^2{ \boldsymbol{k}  } \sqrt{ \textrm{det}( \text{Tr}_\alpha [{ g}^{ (\alpha\sigma), (\alpha\sigma)}( \boldsymbol{k} )] ) },
\end{equation}
which proves the second inequality in Eq.~\eqref{eq:main_ineq}. For the last equality, the above relations for ${ g}^{ (\alpha\sigma), (\alpha\sigma)} (  \boldsymbol{k} )$ combine to give $  { g}^{ (\alpha\uparrow), (\alpha\uparrow)} (  \boldsymbol{k} ) = { g}^{ (\alpha\downarrow), (\alpha\downarrow)} (  \boldsymbol{k} )$, and therefore 
\begin{equation}
  \begin{aligned}
  \sqrt{ \textrm{det} (\textrm{Tr}_b[ g(\boldsymbol{k})] )} &= \sqrt{\text{det}( \text{Tr}_\alpha [ g^{(\alpha\uparrow)(\alpha\uparrow)}(\bs{k})] + \text{Tr}_\alpha [ g^{(\alpha\downarrow)(\alpha\downarrow)}(\bs{k})] ) }
 =2 \sqrt{ \textrm{det} (\text{Tr}_\alpha [g^{(\alpha\sigma), (\alpha\sigma)}(\boldsymbol{k}) ] ) } ,
\end{aligned}
\end{equation} 
which proves Eq.~\eqref{eq:main_ineq}, and hence Eq.~(7) in the manuscript. Lastly, let us recall the definitions $\textrm{vol}_g = \int_{ }^{ } d^2 \boldsymbol{k} \sqrt{ \textrm{det}( \textrm{Tr}_b [ g(\boldsymbol{k})] ) }$ and $\textrm{vol}_{\tilde{g}} = \sqrt{ \int \! d^2{\boldsymbol{k}} \textrm{Tr}_b[ { g}_{ xx}(\boldsymbol{k})  ]  \int \! d^2{\boldsymbol{k}} \textrm{Tr}_b[ { g}_{ yy}(\boldsymbol{k})  ]
- (  \int \! d^2{\boldsymbol{k}} \textrm{Tr}_b[ { g}_{ xy}(\boldsymbol{k})  ] )^2
  } $, which together with Eq.~\eqref{eq:main_ineq} and the already established inequality $ \mathrm{vol}_g \leq \mathrm{vol}_{\tilde{g}}$ yields
\begin{equation}
	 \pi| { \mathcal{C}}_{ \sigma} | \leq \frac{1}{2} \textrm{vol}_g \leq \frac{1}{2} \textrm{vol}_{\tilde{g}}, 
\end{equation}
which is the equation stated in text below Eq.~(7) in the manuscript.

\section{Quantum optics} \label{sec:QO}
In this section, we present the theory of photodetection. 
The photons impinging on the photodetector carry information on the matter source they originate from -- in this case the 
2D material in the cavity.
In order to measure properties of quantum light, we need a device which detects the electric field oscillations.  In a photodetector, the photons excite electrons and by measuring the induced currents, it is possible to deduce the nature of the photon state \cite{Glauber_1963}.  

\subsection{Photodetection theory }

The results in this section closely follow Ref.~\cite{Vogel_2006}. Consider a set of $M$ photodetectors, each comprised of $N$ atoms capable of absorbing light, and assign to each detector a measurement time and let both the detector and measurement time be labeled by ``$l$". We assume that the impinging light is capable of exciting a number $m<N$ of photoelectrons per detector. We define the following operator
\begin{equation} \label{eq:Gamma}
  \hat{ \Gamma}^{(l)} (t_l,\Delta t_l) = \sum^{N}_{i=1} { S}^{ (l,i)} \int_{ t_l}^{ t_l+ \Delta t_l}  d{ \tau} \hat{ E}_{l}^{ (-)'} ( \boldsymbol{r}_{l,i}, \tau) \hat{ E}_{l}^{ (+)'} ( \boldsymbol{r}_{l,i} , \tau) ,
\end{equation}
where $\Delta t_l$ denotes the minimal time resolution of the $l$-th photodetector. $S^{(l,i)}$ is in reality a time dependent factor depending on the dipole matrix elements of the photodetector material, which we assume to be constant in the $l$-th time interval, and independent of the detector. The time dependence of the electric field operators, $\hat{ E}_l^{ (\pm)'} ( \boldsymbol{r}_{l,i} , { \tau})$, is expressed in the Heisenberg picture with respect to the Hamiltonian of the radiation field and evaluated at the position $ \boldsymbol{r}_{l,i}$ of the $i$th atom in the $l$th detector. Crucially, we tacitly assume that a polarization filter is applied prior to the photodetection, allowing us to neglect the vector nature of the electric field. 

In Ref.~\cite{Vogel_2006}, it has been shown that the probability of exciting a set $ \{ m_l | l=1,2,...,M \text{ and } m_l \in [0,N] \}$ of photoelectrons can be computed via 
\begin{equation} \label{eq:P_m}
	P_{ \{ m_l \}} ( \{ t_l, \Delta t_l \} ) = \Bigg\langle{ :  \prod_{l=1}^M \frac{1}{m_l !} [ \hat{\Gamma}^{(l)} (t_l, \Delta t_l)  ]^{m_l} \exp\{ - \hat{\Gamma}^{(l)} (t_l, \Delta t_l)  \} :  }\Bigg\rangle,
\end{equation}
where $  \langle { \dots}\rangle = \textrm{Tr}[ \hat{ \rho} \dots ]  $ is the expectation value in the state $ \hat{ \rho} $ of the incident radiation field, and ``$: :$" is the normal ordering operator which places creation operators to the left and time orders the $\hat{E}_l^{(+)}$ operators by placing operators with later times to the left and $\hat{E}_l^{(-)}$ operators with later times to the right. From Eq.~\eqref{eq:P_m}, we can obtain arbitrary moments of photocounts which are defined as follows 
\begin{equation}
  \begin{aligned}
    \label{eq:HM}
    &\overline{ n( { t}_{ i_1}, \Delta { t}_{ i_1}  ) n( { t}_{ i_2}, \Delta { t}_{ i_2}  )  \cdots n( { t}_{ i_M}, \Delta { t}_{ i_M}  )  }
    &= \sum^{ \infty }_{m_1, m_2, \ldots, m_M = 0} { m}_{ 1} m_2 \cdots m_M P_{ \{ m_l \} }( \{ t_l , \Delta t_l \} ).
  \end{aligned}
\end{equation}
where we have let $N\rightarrow \infty$. Let us point out that the theory allows for higher order moments of a single photodetector by letting $M$ in Eq.~\eqref{eq:P_m} denote the different measurement times at a single detector, provided two photocounting incidents are separated by a time interval larger than $\Delta t$. As an example, $\overline{ n( { t}, \Delta { t} )} $ is the expectation value of the number of photon counts at a detector at time $t$ during the measurement interval $ \Delta t$. 
$\overline{ n( { t}, \Delta { t} )^n} $ can be obtained by generating a distribution of photocounts $ \overline{ n( { t}, \Delta { t} )}$ over several experiments and deducing higher moments from its distribution. $\overline{ n( { t}_{ i_1}, \Delta { t} ) n( { t}_{ i_2}, \Delta { t}  ) } $ involves measuring photocounts at a single detector with a time delay, or  measurements at two different detectors (in the manuscript, we only consider the former case). 
Moments can be related to induced currents as described in Refs.~\cite{Vogel_2006, Loudon_2000} and they find applications in homodyne or heterodyne detection, which we describe in the next section. 

\subsection{Homodyne and heterodyne correlation measurements} \label{sec:homodyne}

We assume that the ensemble of atoms in each photodetector is localized enough that we may approximate Eq.~\eqref{eq:Gamma} as
\begin{equation} \label{eq:gamma_approx}
  \hat{ \Gamma}^{(l)} (t_l,\Delta t_l) \approx \xi^{(l)} \int_{ t_l}^{ t_l + \Delta t_l} \: d{ \tau} \hat{ E }_l^{(-)'} ( \boldsymbol{r}_{l} ,\tau) \hat{ E }_l^{(+)'} ( \boldsymbol{r}_{l} ,\tau) \equiv \xi^{(l)} \int_{ t_l}^{ t_l+ \Delta t_l}  \: d{ \tau} \hat{I}_l^{'} ( \boldsymbol{r}_{l} ,\tau),
\end{equation}
where $ \xi^{(l)} = N  S^{(l)}$ and $S^{(l)}=S^{(l,i)}$ is the factor described previously, which incorporates the dipole matrix elements in the photodetector. $``\boldsymbol{r}_l"$ denotes the average position of the atoms in the $l$th photodetector. With Eqs.~\eqref{eq:HM} and \eqref{eq:gamma_approx}, it can be shown that \cite{Vogel_2006}
\begin{equation} \label{eq:uneq_times}
  \begin{aligned}
    &\overline{ n(t,\Delta t_i) n(t+ \tau, \Delta t_j) } - \overline{ n(t,\Delta t_i) } \cdot \overline{ n(t+ \tau, \Delta t_j) } \\
    &\quad = \xi^{(i)} \Theta ( \Delta t_i - \tau) ( \Delta t_i - \tau) \langle{ \hat{ I}_i^{'} ( \boldsymbol{r}_i ,t)}\rangle \delta_{ij} + 
    \xi^{(i)} \xi^{(j)} \Delta t_i \Delta t_j \langle{ : \Delta \hat{ I}_i^{'} ( \boldsymbol{r}_i , t) \Delta \hat{ I}_j^{'} ( \boldsymbol{r}_j , t+ \tau) : }\rangle ,
  \end{aligned}
\end{equation}
where $i,j=1,2$ denotes the output port/detector. 
$\Theta(t)$ is the Heaviside step function and we have approximated $\int_{ t}^{ t+ \Delta t}  \: d{ \tau} \hat{I}_i^{'} ( \boldsymbol{r}_i ,\tau) \approx \Delta t_i \hat{I}_i^{'} ( \boldsymbol{r}_i ,t)$. The first term reflects the classical shot noise term which is indicative of coherent light having a Poissonian photocounting statistics for $\tau=0$ \cite{Vogel_2006}. The second term is purely quantum mechanical in nature and will be our main focus. 

One can relate the operators of photons entering the beam splitter to the photon operators 
exiting the beam splitter 
by means of a unitary scattering matrix as follows 
\begin{equation}
  \begin{pmatrix}
    \hat{b}_{1}^{'} \\
    \hat{b}_{2}^{'} 
  \end{pmatrix} 
  = \begin{pmatrix}
    \mathcal{T} & \mathcal{R} \\
    - { \mathcal{R}}^{ *} & { \mathcal{T}}^{ *} \\
  \end{pmatrix} 
  \begin{pmatrix}
    \hat{ b}_{ 1} \\
    \hat{ b}_{ L} 
  \end{pmatrix} ,
\end{equation}
where $ \hat{b}_1, \hat{b}_L $ refers to the ingoing fields and $ \hat{b}_1^{'}, \hat{b}_2^{'} $ to the outgoing ones. $\mathcal{T}, \mathcal{R}$, which satisfy $| \mathcal{T}|^2 + | \mathcal{R}|^2=1$, are transmission and reflection coefficients, respectively. Although $\mathcal{T}, \mathcal{R}$ are frequency dependent objects, we approximate them as constant over the frequency range of the radiation incident on the beam-splitter. If we assume equal field strength, $\mathcal{E}$, of both outgoing electric fields, we have 
\begin{equation}
\begin{aligned}
	\hat{E}_1^{(+)' }(\bs{r}_1, t) &= i\mathcal{E}\hat{b}_1^{'} (t) e^{i\bs{k}_1 \cdot \bs{r}_1}= i\mathcal{E} e^{i\bs{k}_1 \cdot \bs{r}_1} [ \mathcal{T} \hat{b}_1(t) + \mathcal{R} \hat{b}_L(t) ] ,\\
	\hat{E}_2^{(+)' }(\bs{r}_2, t) &= i\mathcal{E} \hat{b}_2^{'} (t) e^{i\bs{k}_2 \cdot \bs{r}_2}= i\mathcal{E} e^{i\bs{k}_2 \cdot \bs{r}_2} [ -\mathcal{R}^* \hat{b}_1(t) + \mathcal{T}^* \hat{b}_L(t) ] .
\end{aligned}
\end{equation}
Thus, upon defining $\hat{I}_i^{(')} ( \boldsymbol{r}_i ,t) =\hat{ E }_i^{(-)(')} ( \boldsymbol{r}_i ,t) \hat{ E }_i^{(+)(')} ( \boldsymbol{r}_i ,t)$, the measured intensities at the first and second output ports are
\begin{equation}
    \label{eq:shot_noise}
    \begin{aligned}
    \hat{ I}_1^{'} ( \boldsymbol{r}_1,t) &= | \mathcal{T}|^2 \hat{ I}_1( \boldsymbol{r}_1 , t) + | \mathcal{R}|^2 { \hat{ I} }_{ L}(\bs{r}_1) + \mathcal{T}^* \mathcal{R} { \hat{ {{E}}} }_1^{ (-)} (\bs{r}_1, t) { \hat{ {E}} }_{ L}^{ (+)} (\bs{r}_1,t) + \mathcal{R}^* \mathcal{T} { \hat{ E }_{ L}^{ (-)} (\bs{r}_1,t) { \hat{ E}} }_1^{ (+)} (\bs{r}_1,t), \\
    \hat{ I}_2^{'} (\boldsymbol{r}_2,t) &= | \mathcal{T}|^2 \hat{ I}_1(\boldsymbol{r}_2 , t) + | \mathcal{R}|^2 { \hat{ I} }_{ L}(\bs{r}_2) - \mathcal{T}^* \mathcal{R} { \hat{E} }_1^{ (-)} (\bs{r}_2, t) { \hat{E} }_{ L}^{ (+)} (\bs{r}_2, t) - \mathcal{R}^* \mathcal{T} { \hat{E} }_{ L}^{ (-)} (\bs{r}_2, t) { \hat{ E} }_1^{ (+)} (\bs{r}_2, t),
\end{aligned}
\end{equation}
where $\hat{E}_{\alpha}^{ (+)}(\bs{r}_{1(2)},t )  = i {\mathcal{E}} \hat{b}_\alpha^{} (t) e^{ i \boldsymbol{k}_{1(2)} \cdot \boldsymbol{r}_{1(2)}}$, $\hat{E}^{(-)} = [\hat{E}^{(+)}]^\dagger$ with $\alpha=1,L$ and where the absence of time dependence in $\hat{I}_L( \bs{r}_{1(2)})$ is due to monochromatic light in the input port of the local oscillator. Now, we assume the expectation values in Eq.~\eqref{eq:uneq_times} to be taken over the state $\hat{\rho} = \hat{\rho}_{ \textrm{out}} \otimes |{\alpha_L} \rangle  \langle{ \alpha_L}| $, where $\rho_{\text{out}}$ is the density matrix of the photons exiting the cavity and $ |{ \alpha_L} \rangle $ is a coherent state satisfying $ \hat{b}_L |{ \alpha_L}\rangle = | \alpha_L | e^{ i \phi }  |{ \alpha_L} \rangle $. Evaluating Eq.~\eqref{eq:uneq_times} while assuming a strong local oscillator field ($|\alpha_L|$ large with $\overline{n(t,\Delta t_i)} \approx \xi^{(i)} \Delta t_i | \mathcal{R}|^2 | \mathcal{E}|^2 | \alpha_L |^2 $), it can be shown that for measurements involving a \emph{single detector} ($i=j=1,2$ in Eq.~\eqref{eq:uneq_times}) 
\begin{equation} \label{eq:final_res}
  \begin{aligned}
    &\frac{ \overline{ n(t, \Delta t_{1(2)}) n(t+\tau, \Delta t_{1(2)})  } - \overline{n(t,\Delta t_{1(2)}) } \cdot  \overline{ n(t+ \tau, \Delta t_{1(2)}) } }{ \overline{ n(t, \Delta t_{1(2)}) }} \\
    & \approx  \Theta ( \Delta t_{1(2)} - \tau) \frac{ \Delta t_{1(2)} - \tau}{ \Delta t_{1(2)}} \pm \xi^{( 1(2) )} \Delta t_{1(2)} | \mathcal{T}|^2 \langle { : \Delta \hat{ E}_{1} (\bs{r}_{1(2)}, t , \phi ) \Delta \hat{ E}_{1} ( \bs{r}_{1(2)}, t+ \tau, \phi )   :}\rangle ,
  \end{aligned}
\end{equation}
where, due to the nature of the coherent state, the expectation value in Eq.~\eqref{eq:final_res} is defined as $\langle \dots \rangle  = \text{Tr}[\rho_{\text{out}} \dots ]$. We have also defined 
\begin{equation}
  \hat{ E}_1 (\bs{r}_{1(2)}, t, \phi ) = \hat{ E }_1^{ (+)} (\bs{r}_{1(2)}, t) e^{ i \omega_L t} e^{-i \phi }  + \text{H.c.}, 
\end{equation}
$\Delta \hat{{E}}_1(\bs{r}_{1(2)},t,\phi) = \hat{{E}}_1(\bs{r}_{1(2)},t,\phi) - \langle : \hat{{E}}_1(\bs{r}_{1(2)},t,\phi) : \rangle$ and we assumed that $\mathcal{T}=\mathcal{R}$. The factor $e^{ i \omega_L t}$ cancels any time dependence of the form $e^{ -i \omega_L t} $ in $ \hat{ {E}}_1^{(+)} (\bs{r}_{1(2)},t)$. When this cancellation is exact, it is referred to as homodyne detection, otherwise it is referred to as heterodyne detection. Finally, we mention a procedure called \emph{balanced} homodyne/heterodyne detection which involves subtracting the signals at the different detectors (see Eq.~\eqref{eq:final_res}). This procedure can be advantageous since a large local oscillator field strength is not necessary to arrive at the last terms on the RHS of Eq.~\eqref{eq:final_res}. In the following sections, we apply Eq.~\eqref{eq:final_res} to the emitted field from the cavity.

\section{Input-output formalism}

\subsection{Intracavity and environment photons}

In order to describe the fields emitted from the cavity, we extend our system Hamiltonian to include contributions from the environment as well as a system-environment coupling. The full Hamiltonian reads 
\begin{equation} \label{eq:HSE}
  \hat{ H}_\text{SE} = \hat{H}_{\textrm{cav}} + \int d\omega \hbar \omega  \hat{ b}^{ \dagger } ( \omega) \hat{ b} ( \omega)   +\hbar \int d\omega [  \mathcal{M} ( \omega)  \hat{ a}^{ \dagger } \hat{ b} ( \omega) + h.c. ]   ,
\end{equation}
where $\hat{H}_{\textrm{cav}}$ is the intra-cavity Hamiltonian as defined in the manuscript, $\hat{a}$ refers to the main intracavity mode operator, $\{ \hat{b}(\omega) \}$ is the continuum of environment photon modes which couple to the cavity, while $\mathcal{M}(\omega)$ is the cavity-environment coupling. From Ref.~\cite{Viviescas_2003}, we find from the Heisenberg equation of motion that
\begin{equation}\label{eq:in_out}
  \hat{ b}_{ }^{ \text{out}}(t) - \hat{ b}_{ }^{ \text{in}} (t) = - \frac{ i}{ 2 \pi}  \int_{ -\infty}^{ \infty} d\omega \int_{ t_0}^{ t_1} dt' { \mathcal{M}}_{}^{*} ( \omega) e^{ -i \omega(t-t')} \hat{ a} (t') \approx -i { \mathcal{M}}_{ }^{ *} \hat{ a} (t)  ,
\end{equation}
where $\hat{ b}_{ }^{ \textrm{out}} (t) = \frac{ 1}{ 2 \pi} \int_{ - \infty }^{ \infty } e^{ -i \omega(t-t_1)} \hat{ b}_{ } ( \omega,t_1) \: d{ \omega} $ and $\hat{ b}_{ }^{ \textrm{in}} (t) = \frac{ 1}{ 2 \pi} \int_{ - \infty }^{ \infty } e^{ -i \omega(t-t_0)} \hat{ b}_{} ( \omega,t_0) \: d{ \omega}$. Here, $\hat{b}(\omega,t)$ are the operators $\hat{b}(\omega)$ in the Heisenberg picture with respect to Eq.~\eqref{eq:HSE} and we assumed $ \mathcal{M}( \omega)$ to be frequency independent as in Ref.~\cite{Walls_2007}. Equation~\eqref{eq:in_out} is an important relation which enables us to relate intra-cavity photon correlators to correlators of output fields.

\subsection{Correlation functions} 
Using Eq.~\eqref{eq:in_out}, let us assume a coherent state input. This essentially means that the state of the system described by the cavity plus its environment can be described by something akin to a product state, $\hat{\rho}_{SE} \approx | \alpha_{\text{in}} \rangle \langle \alpha_{\text{in}} | \otimes \hat{\rho}_{\text{cav}} \otimes \hat{\rho}_{\text{out}}$, where $| \alpha_{\text{in}} \rangle $ is the coherent state of the input field (not to be confused with the coherent state of the previous section). By Eq.~\eqref{eq:in_out}, we can write \cite{Walls_2007}
\begin{equation} \label{eq:in_out_corr}
  \langle \hat{ b}_{ 1}^ \dagger (t), \hat{ b}_{ 1} (t') \rangle \equiv \langle{ { \hat{ b} }^{ \textrm{out}, \dagger } (t), { \hat{ b} }^{ \textrm{out}} (t')} \rangle \approx |{ \mathcal{M}}|^2 \langle{ { \hat{ a} }^{ \dagger } (t), { \hat{ a} } (t')} \rangle,
\end{equation}
where $ \langle{ U,V} \rangle= \langle{ UV}\rangle - \langle{ U}\rangle \langle{ V}\rangle$ and we have defined $\hat{b}^{\text{out}}=\hat{b}_1$ in order to connect to the theory in Sec.~\ref{sec:homodyne}. 
With the help of Eq.~\eqref{eq:in_out_corr}, we can write the electric field correlator at a \emph{single detector} derived in Eq.~\eqref{eq:final_res} in terms of the intra-cavity correlator (to be derived in Sec.~\ref{sec:LM}). Due to the fact that expressions involving oscillations in $ t + t'$ can be removed by a suitable time averaging procedure, we collect terms depending on $ t_{\text{rel}} \equiv t-t'$ as follows
\begin{equation} \label{eq:rel_to_corr}
\begin{aligned}
	\langle { : \Delta \hat{ E}_{1} (\bs{r}_{1}, t , \phi ) \Delta \hat{ E}_{1} ( \bs{r}_{1}, t', \phi )   :}\rangle &= e^{ - i \omega_L t_{\text{rel}}} \langle{ \Delta \hat{{E}}_1^{(-)}(\bs{r}_1, t) \Delta \hat{{E}}_1^{(+)}(\bs{r}_1, t') }\rangle +e^{  i \omega_L t_{\text{rel}}} \langle{ \Delta \hat{{E}}_1^{(-)}(\bs{r}_1, t') \Delta \hat{{E}}_1^{(+)}(\bs{r}_1, t) }\rangle \\
	& \approx | \mathcal{M}|^2 |\mathcal{E} |^2 \{   e^{ - i \omega_L t_{\text{rel}}} \langle{ { \hat{ a} }^{ \dagger } (t) ,\hat{ a} (t')}\rangle + e^{  i \omega_L t_{\text{rel}}}  \langle{{ \hat{ a} }^{ \dagger } (t' ) ,\hat{ a} (t)  }\rangle \}.
\end{aligned}
\end{equation}
This expression leaves us with the task of computing the intra-cavity correlator, $\langle{ { \hat{ a} }^{ \dagger } (t) ,\hat{ a} (t')}\rangle $, which will be the topic of the next section.

\section{Intra-cavity Light-matter coupling } \label{sec:LM}

\subsection{Correlators calculated in the Heisenberg picture}
To perform calculations involving the photon correlation functions, as in Eq.~\eqref{eq:rel_to_corr}, we rely on decomposing the intracavity light-matter Hamiltonian into an interacting and a non-interacting part. The Gell-Mann Low theorem \cite{Stefanucci_2013} states that correlators computed over an interacting Hamiltonian
\begin{equation}
  \hat{ H}_{\eta} (t) = \hat{ H}_0 + \zeta(t) { \hat{ H} }_{ I} , 
\end{equation}
where $\zeta(t) \equiv \theta(t-t_0) + e^{ - \eta |t-t_0|} \theta (t_0 - t)$, can be computed over the non-interacting density matrix of $\hat{H}_0$, $\hat{\rho}_0$, by adiabatically switching on the interaction. $\eta$ is a positive infinitesimal used to slowly turn on $\hat{H}_{I}$ from an initial time, $t_1<t_0$. This enables us to express the two-point correlator in the following way \cite{Stefanucci_2013},
\begin{equation} \label{eq:correlator}
  \langle{ \hat{ \mathcal{O}} (t) \hat{ \mathcal{O}} (t')} \rangle = \textrm{Tr}[ { \rho}_{ 0} \hat{U}_{ \eta }(t_1 ,t) \hat{ \mathcal{O}} \hat{ U}_{ \eta } (t, t_1 ) { \hat{U}}_{ \eta } (t_1 , t') \hat{ \mathcal{O}} { \hat{U}}_{ \eta } (t', t_1 ) ],
\end{equation}
where we denote the time evolution operator of the system by $ \hat{U}_{\eta}(t_2, t_1) = \hat{\mathcal{T}} \text{exp}\{ -\frac{i}{\hbar} \int_{t_1}^{t_2} dt \hat{H}_{\eta}(t) \}$ and $\hat{\mathcal{T}}$ is the time ordering operator. The time dependence of operators is computed from $\partial_{t} \hat{\mathcal{O}}(t) = (i/\hbar) [ \hat{H}_\eta(t), \hat{\mathcal{O}}(t)] $. If we define 
\begin{equation}
	\hat{ H}_{0} = \sum^{}_{\boldsymbol{k},\alpha} { \epsilon}_{ \alpha}(\boldsymbol{k}) \hat{ c}_{ \boldsymbol{k}, \alpha}^{ \dagger } \hat{ c}_{ \boldsymbol{k}, \alpha} + \hbar \Omega \hat{a}^\dagger \hat{a}, 
\end{equation}
and $\hat{H}_I = \hat{H}_{\text{cav}} - \hat{H}_0$ with $ \hat{H}_{\text{cav}}$ defined in the manuscript, we find that 
\begin{equation} \label{eq:EOM}
  \begin{aligned}
  \partial_{t} \hat{ a} (t) &\approx -i \tilde{ \Omega}(t) \hat{ a} (t) - \zeta(t) \frac{ i}{ \hbar } \frac{ \lambda^2 q^2}{ m} {{ \boldsymbol{f} }^{ *}}^{2} { \hat{ a} }^{ \dagger }(t) 
  + \zeta(t)  \frac{ i \lambda q}{ \hbar^2} { \boldsymbol{f} }^{ *} \sum^{}_{ \boldsymbol{k}} \sum_{ \alpha, \beta=1}^N e^{  i\epsilon_{\alpha\beta}(\boldsymbol{k}) t } { \hat{ c} }_{ \boldsymbol{k} ,\alpha}^{ \dagger } (t_1) \langle u_{\bs{k},\alpha} | \nabla_{\bs{k}} \hat{\mathcal{H}}(\bs{k}) | u_{\bs{k},\beta} \rangle  { \hat{ c} }_{ \boldsymbol{k} , \beta}(t_1)  , 
\end{aligned}
\end{equation}
where $\tilde{ \Omega}(t) = \Omega + \zeta(t) \frac{ \lambda^2 q^2}{ m \hbar } \boldsymbol{f} \cdot \boldsymbol{f}^* $ and where we have used that to leading order, $ \hat{c}_{\boldsymbol{k}\alpha}(t) \approx e^{ -\frac{i}{\hbar}  \epsilon_\alpha (\bs{k}) (t-t_1)} \hat{c}_{\boldsymbol{k}, \alpha}(t_1)$. The orbital space model, defined by $\hat{\mathcal{H}}(\bs{k})$, has been incorporated into Eq.~\eqref{eq:EOM} by approximating the velocity matrix elements as
\begin{equation}
	\langle \psi_{\bs{k},\alpha} | \hat{\bs{p}} | \psi_{\bs{k},\beta} \rangle \approx \frac{m}{\hbar} \langle u_{\bs{k},\alpha} | \nabla_{\bs{k}} \hat{\mathcal{H}}(\bs{k}) | u_{\bs{k},\beta} \rangle.
\end{equation}
The result is a coupled set of equations for $ \hat{ a} $ and $ { \hat{ a} }^{ \dagger } $ and it can be easily verified that a solution to Eq.~\eqref{eq:EOM} is
\begin{equation} \label{eq:a_TD}
  \begin{aligned}
    \hat{ a} (t) =& \, e^{ -i \int_{t_1}^t dt' \tilde{ \Omega}(t') } \Big\{  \hat{ a} (t_1) +i \int_{ t_1 }^{ t} d{t' }  \frac{ \lambda q}{ \hbar^2} \boldsymbol{f}^* \sum^{}_{ \boldsymbol{k} \alpha\beta} { \hat{ c} }_{ \boldsymbol{k} \alpha}^{ \dagger }(t_1)  \langle u_{\bs{k},\alpha} | \nabla_{\bs{k}} \hat{\mathcal{H}}(\bs{k}) | u_{\bs{k},\beta} \rangle \hat{ c}_{ \boldsymbol{k} \beta}(t_1) e^{ i \int_{t_1}^{t'} dt'' \tilde{ \Omega}(t'')} e^{  i\epsilon_{\alpha\beta}(\boldsymbol{k}) t'} \zeta(t')     \\
    &- \int_{t_1}^t dt' \zeta(t') \frac{i}{\hbar}\frac{ q^2 \lambda^2}{ m} { \boldsymbol{f} }^{ *,2} e^{ i \int_{t_1}^{t'} dt'' \tilde{\Omega} (t'') } { \hat{ a} }^{ \dagger } (t')  \Big\} , \\
  \end{aligned}
\end{equation}
which can be further simplified: The weak coupling assumption
allows us to drop the last term, and since $ \int_{t_1}^{t'} dt'' \tilde{ \Omega}(t'') = \tilde{\Omega} (t'-t_1) + \mathcal{O}(\eta) $, the value of the integral for small $ \eta $ is \footnote{In order to calculate the second term in Eq.~\eqref{eq:a_TD}, note that 
\begin{equation}
\begin{aligned}
	\int_{t_1}^t \zeta(t') dt' &= (t-t_1) + \int_{t_1}^{t_0} e^{ -\eta (t_0 - t')} dt' = (t-t_0)  - \frac{1}{\eta}(1 - e^{-\eta(t_0 - t_1)} ) = (t-t_1) + \mathcal{O}(\eta).
\end{aligned}
\end{equation}}
\begin{equation}
  \int_{ t_1 }^{ t} dt' e^{ i ( \tilde{\Omega} +  { \epsilon}_{ \alpha\beta}(\bs{k}) ) t'} [(\theta(t'-t_0) + e^{ - \eta |t'-t_0|} \theta (t_0 - t'))  ]  \approx \frac{ e^{ i( \tilde{\Omega} + \epsilon_{\alpha\beta}(\bs{k}) )t } }{ i ( \tilde{\Omega} + \epsilon_{\alpha\beta}(\bs{k}) )} ,
\end{equation}
where we have taken $ (t_0 -t_1) \rightarrow \infty$, set $t_1 =0$ and made the approximation $\tilde{\Omega}(t) \approx \tilde{\Omega}\equiv \Omega + \frac{ \lambda^2 q^2}{ m \hbar } \boldsymbol{f} \cdot \boldsymbol{f}^* $ for all $t$ (weak coupling approximation). With these approximations, we can finally write 
\begin{equation}
\begin{aligned}
	\hat{a}(t) &\approx e^{-i \tilde{\Omega} t} \Bigg\{ \hat{a}(0) + i \frac{\lambda q}{\hbar^2 } \bs{f}^* \cdot \sum_{\bs{k}, \alpha\beta } \frac{e^{i(\tilde{\Omega}  + \epsilon_{\alpha\beta}(\bs{k}))t}}{ i(\tilde{\Omega} + \epsilon_{\alpha\beta}(\bs{k}) )  } \hat{c}_{\bs{k},\alpha}^\dagger (0)  \langle u_{\bs{k},\alpha} | \nabla_{\bs{k}} \hat{\mathcal{H}}(\bs{k}) | u_{\bs{k},\beta} \rangle \hat{c}_{\bs{k},\beta}(0)  \Bigg\}.
\end{aligned}
\end{equation}

 Equation~\eqref{eq:correlator} can now be computed by specifying $\rho_0$ as a product state between electron and photon states, $\rho_0 = \rho_{\text{el}} \otimes | 0 \rangle \langle 0 |$, where $\rho_{\text{el}} = \prod_{\alpha=1}^{M} \prod_{\boldsymbol{k} \in \text{BZ}} \otimes | u_{\boldsymbol{k},\alpha} \rangle \langle u_{\boldsymbol{k}, \alpha} |$. This leads to the following correlator 
\begin{equation}
  \begin{aligned}
    \label{eq:final_corr}
    \langle{ { \hat{ a} }^{ \dagger } (t), \hat{ a} (t')} \rangle \equiv \langle{ { \hat{ a} }^{ \dagger } (t) \hat{ a} (t')}\rangle - \langle{ \hat{ a}^{ \dagger } (t)}\rangle \langle{ \hat{ a} (t')}\rangle 
    \approx ( \frac{ q\lambda}{ \hbar } )^2 \sum^{}_{\bs{k}} \sum_{\alpha=1}^{M} \sum_{\beta=M+1}^{N} e^{ - i  { \epsilon}_{ \alpha\beta}(\boldsymbol{k}) t_{\text{rel}}} { f}^{ \mu} { f}^{ \nu *}  { \mathcal{A}}^{ \alpha, \beta}_\mu (\boldsymbol{k}) { \mathcal{A}}^{ \beta, \alpha}_\nu (\boldsymbol{k})  ,
  \end{aligned}
\end{equation}
where we used Eq. \eqref{eq:vel_to_Berry} in the last line, while assuming that $ \tilde{\Omega} \ll  \tfrac{1}{\hbar}| \epsilon_{\alpha}(\boldsymbol{k}) - \epsilon_{\beta}(\boldsymbol{k})|$, which is valid away from band-touchings and for weak light-matter coupling.

\subsection{Final result}
Having computed the intra-cavity correlator in Eq.~\eqref{eq:final_corr} and related it to an output correlator through Eq.~\eqref{eq:rel_to_corr}, we arrive at the following expression
\begin{equation} \label{eq:rel_to_corr_final}
\begin{aligned}
	\langle { : \Delta \hat{ E}_{1} (\bs{r}_{1}, t , \phi ) \Delta \hat{ E}_{1} ( \bs{r}_{1}, t', \phi )   :}\rangle
	\approx | \mathcal{M}|^2 |\mathcal{E} |^2  ( \frac{ q\lambda}{\hbar } )^2  \sum^{}_{\bs{k}} \sum_{\alpha=1}^{M} \sum_{\beta=M+1}^{N} \big[ e^{ i ( { \epsilon}_{ \beta\alpha}(\boldsymbol{k}) - \omega_L) t_{\text{rel}}} + \text{h.c.} \big] { f}^{ \mu} { f}^{ \nu *} { \mathcal{A}}^{ \alpha, \beta}_\mu(\boldsymbol{k}) { \mathcal{A}}^{ \beta, \alpha}_\nu(\boldsymbol{k}).
\end{aligned}
\end{equation}
By Eq.~\eqref{eq:final_res} for the first detector, the final result for the photodetection theory therefore becomes
\begin{equation}\label{eq:PD_result}
\begin{aligned} 
	   &\frac{ \overline{ n(t, \Delta t) n(t', \Delta t)  } - \overline{n(t,\Delta t) } \cdot  \overline{ n(t', \Delta t) } }{ \overline{ n(t, \Delta t) }} \\
	   & \quad \approx \xi | \mathcal{T}|^2 (\Delta t)  |\mathcal{M}|^2 |\mathcal{E}|^2  \Big( \frac{q\lambda}{\hbar} \Big)^2 \sum^{}_{\bs{k}} \sum_{\alpha=1}^{M} \sum_{\beta=M+1}^{N} \big[ e^{ i ( { \epsilon}_{ \beta\alpha}(\boldsymbol{k}) - \omega_L) t_{\text{rel}}} + \text{h.c.} \big] { f}^{ \mu} { f}^{ \nu *} { \mathcal{A}}^{ \alpha, \beta}_\mu(\boldsymbol{k}) { \mathcal{A}}^{ \beta, \alpha}_\nu(\boldsymbol{k})   \\
\end{aligned}
\end{equation}
if we assume $|t'-t|> \Delta t$. This is Eq.~(3) in the manuscript with $\mathcal{D} = \xi | \mathcal{T}|^2 (\Delta t)  |\mathcal{M}|^2 |\mathcal{E}|^2 ( \frac{q\lambda}{\hbar} )^2$.

\bibliography{bibliography.bib}